# Language-specific Tonal Features Drive Speaker-Listener Neural Synchronization

Chen Hong[a,b], Xiangbin Teng[b,c], Yu Li[d], Shen-Mou Hsu[e], Feng-Ming Tsao[f], Patrick C. M. Wong[a,b], Gangyi Feng[a,b *]

a. Department of Linguistics and Modern Languages, The Chinese University of Hong Kong, Shatin, N.T., Hong Kong SAR, China
b. Brain and Mind Institute, The Chinese University of Hong Kong, Shatin, N.T., Hong Kong SAR, China
c. Department of Psychology, The Chinese University of Hong Kong, Shatin, N.T., Hong Kong SAR, China
d. Applied Psychology Programme, BNU-HKBU United International College, China
e. Imaging Center for Integrated Body, Mind and Culture Research, National Taiwan University, Taipei, Taiwan, China
f. Department of Psychology, National Taiwan University, Taipei, Taiwan, China

***Corresponding author:**

Gangyi Feng, g.feng@cuhk.edu.hk

Brain and Mind Institute
Department of Linguistics and Modern Languages
The Chinese University of Hong Kong, Shatin, N.T., Hong Kong SAR, China

**Abstract**

Verbal communication transmits information across diverse linguistic levels, with neural synchronization (NS) between speakers and listeners emerging as a putative mechanism underlying successful exchange. However, the specific speech features driving this synchronization and how language-specific versus universal characteristics facilitate information transfer remain poorly understood. We developed a novel content-based interbrain encoding model to disentangle the contributions of acoustic and linguistic features to speaker-listener NS during Mandarin storytelling and listening, as measured via magnetoencephalography (MEG). Results revealed robust NS throughout frontotemporal-parietal networks with systematic time lags between speech production and perception. Crucially, suprasegmental lexical tone features (tone categories, pitch height, and pitch contour), essential for lexical meaning in Mandarin, contributed more significantly to NS than either acoustic elements or universal segmental units (consonants and vowels). These tonal features generated distinctive spatiotemporal NS patterns, creating language-specific neural “communication channels” that facilitated efficient representation sharing between interlocutors. Furthermore, the strength and patterns of NS driven by these language-specific features predicted communication success. These findings demonstrate the neural mechanisms underlying shared representations during verbal exchange and highlight how language-specific features can shape neural coupling to optimize information transfer during human communication.

## Introduction

Communication with language is a unique human ability that enables the efficient transmission of complex thoughts and emotions through speech. It plays a fundamental role in facilitating cooperation and knowledge sharing in human societies. This sophisticated capability relies on coordinated neural processes involving distributed networks between brains that collectively support speech production and perception (Hertrich et al., 2020; Hickok, 2012; Hickok & Poeppel, 2007; Price, 2012; Rauschecker & Scott, 2009; Simonyan & Fuertinger, 2015; Stephens et al., 2010). Understanding how these networks support the dynamic exchange of linguistic information during communication is a central question across domains of cognitive neuroscience.

The neural mechanisms underlying speech production and perception have contributed greatly to our understanding of how humans process speech. Speech perception is primarily linked to the auditory pathway, involving the superior temporal gyrus (STG) and Heschl's gyrus (HG), which process acoustic signals and extract speech cues for comprehension (Bhaya-Grossman & Chang, 2022; Binder et al., 2000; DeWitt & Rauschecker, 2012; Friederici, 2012; Friederici et al., 2003; Giraud & Poeppel, 2012; Nuñez et al., 2020; Rauschecker & Scott, 2009). In contrast, speech production engages sensorimotor regions like the precentral and postcentral gyri, responsible for controlling articulatory processes (Bouchard et al., 2013; Chartier et al., 2018; Correia et al., 2020; Dastolfo-Hromack et al., 2022; Hickok, 2012). The frontotemporal-parietal regions such as the anterior temporal lobe (ATL), middle temporal gyrus (MTG), inferior frontal gyrus (IFG), and inferior parietal lobe (IPL) play roles in higher-level processes common across speech perception, production planning, and various aspects of speech control (Flinker et al., 2015; Sheng et al., 2019; Silbert et al., 2014; Tourville & Guenther, 2011; Tremblay et al., 2019). Recent findings highlight overlapping neural circuits for speech production and perception, implying a common network for language communication (Evans & Davis, 2015; Fairs et al., 2021; Pulvermüller & Fadiga, 2010; Wilson et al., 2004). This overlap occurs not only in the anatomical areas but also in the neural representations of shared content. This can help quickly map the shared representations between the speaker's production and the listener's perception processes, which is essential for effective verbal communication.

Although single-brain neuroimaging studies have offered significant insights into the neural mechanisms of speech processing, they fail to fully represent the neural interactions occurring between brains during communication between interlocutors. (Dumas et al., 2010; Hasson et al., 2012; Jiang et al., 2021; Kingsbury & Hong, 2020). Recent developments in neuroimaging methods, including hyperscanning, allow for the capturing of neural activities from multiple individuals during communication (Zhou et al., 2024; Zhou & Wong, 2024). This has uncovered interbrain neural synchronization (NS) as a possible neural mechanism for successful verbal information exchange (Jiang et al., 2012; Liu et al., 2021; Stephens et al., 2010). NS refers to aligned neural activity between individuals, particularly in regions supporting speech production and perception (Z. Li et al., 2021; Liu et al., 2020; Stephens et al., 2010). Stronger speaker-listener NS often correlates with better comprehension and shared understanding (Hasson et al., 2012; Liu et al., 2021; Pérez et al., 2017).

Accumulating evidence indicates that various factors, including predictable language context (Dikker et al., 2014), shared attention (Pérez et al., 2019), emotional engagement (Smirnov et al., 2019), and interactive contexts (Koul et al., 2023; Pan et al., 2022) can modulate NS. There is also interest in acoustic influences on interbrain similarity, such as the role of speech envelope or pitch (Luo & Poeppel, 2007; Peelle et al., 2013). However, the specific linguistic units and speech features that underpin NS remain poorly understood, especially regarding language-specific units (e.g., lexical pitch and tone categories in tone languages) versus language-universal units (e.g., phonemes, syllables, prosodic patterns).

In tone languages like Mandarin Chinese, pitch contours are integral to lexical meaning, serving as language-specific speech cues that distinguish one word from another (Duanmu, 2007; Howie, 1976). For example, the syllable "ma" may refer to "mother," "hemp," "horse," or "scold" depending on its pitch pattern. Processing these pitch variations relies on robust representations of pitch height and pitch change in frontotemporal-parietal networks (Feng, Gan, Llanos, et al., 2021; Feng et al., 2018; Feng, Gan, Yi, et al., 2021; Feng et al., 2019; Gandour et al., 2000; Y. Li et al., 2021). Such findings suggest that tone-relevant neural computations and possibly their interbrain correlates might be distinct, reflecting specialized language-specific processing demands. Investigating how language-specific (e.g., tonal) and language-universal (e.g., segmental) units affect interbrain NS could provide important insights into both shared and language-dependent aspects of neural communication mechanisms.

In this study, we examined the extent to which different speech features and sublexical units (e.g., phonemes, lexical tones) shape speaker-listener NS in Mandarin. We focus on both language-specific and language-universal units to understand their distinct roles in driving NS. We hypothesize that effective communication relies on shared neural representations of linguistic and communicative content, with different linguistic units utilizing distinct NS patterns and mechanisms. Considering the critical role of tones in Mandarin, where pitch variations signify meaning (Duanmu, 2007; Howie, 1976), we anticipate that tone categories and tone-related speech features, such as pitch height and pitch change, would modulate NS more significantly than non-tonal segmental features like consonants and vowels, which tend to be more universally applicable across languages. We examine the spatiotemporal patterns of NS to reveal how different linguistic units and speech features utilize distinct NS patterns as neural "communication channels" to support the sharing of linguistic representations (Friston, 2005; Giraud & Poeppel, 2012) and the directionality of information flow in communication between speakers and listeners across regions and hemispheres (Brodbeck et al., 2018; Sahin et al., 2009).

Testing our hypotheses required a neuroimaging method that offers both high spatial and temporal precision, coupled with a novel modeling framework capable of incorporating linguistic content into analyses of joint neural activity. Consequently, we conducted a pseudo-hyperscanning magnetoencephalography (MEG) study using a storytelling-listening paradigm. A selected native Mandarin speaker narrated eight stories after reading their scripts, and a group of native Mandarin-speaking listeners were subsequently scanned while listening to these recorded narratives (see Figure 1A). We implemented a new content-based neural encoding modeling approach that uses temporal response functions (TRFs) to predict listeners' neural responses based on the speaker's neural activity to various speech features and linguistic units (see Figure 1B). By quantifying the predictive contribution of language-universal units (e.g., phonemes, syllables) and language-specific units (e.g., lexical tones, pitch features) to speaker-listener neural predictions, we assessed their respective roles in driving NS. Higher prediction accuracy indicates stronger NS, reflecting shared neural representations promoted by these linguistic features. This approach enabled us to identify interbrain patterns of NS across regions and time-lag windows, revealing the distinct synchronization mechanisms that facilitate communication with universal and language-specific features. Finally, we used these language-universal and language-specific NS patterns to predict individual communication outcomes, verifying the behavioral importance of the identified spatiotemporal NS signatures.

## Results

### Story comprehension performance

Six speakers spontaneously told eight stories one by one after reading the corresponding text scripts while undergoing MEG scanning. We recorded their stories (~56 min) and MEG data and then selected one speaker's audio recordings based on the fluency of the storytelling and the quality of the MEG data (Fig. S1A, Supplementary). The audio recordings were then played back to 22 listeners. To ensure that the listeners paid attention to the stories, we asked them to answer 80 true-or-false questions related to the content of the stories (10 questions per story). The average accuracy was 92.2% (standard deviation [SD] = 6.5%; range: 72.5%-98.8%), significantly above the chance level ($t_{21} = 34.0$, $p < 0.001$).

### Spatial patterns of speaker-listener neural synchronization (NS)

We quantified the speaker-listener NS using the temporal response function (TRF)-based encoding modeling approach(Crosse et al., 2016; J. Li et al., 2023). We modified the original TRF-based approach for interbrain synchronization modeling (see Fig. 1 for the graphical illustration), which enables us to examine NS by using the speaker's neural activities to predict listeners' neural activities and assess to what extent the speech and linguistic features contribute to the inter-brain predictions.

To assess the overall speaker-listener NS, we first used the speaker's MEG responses to predict listeners' MEG responses without including feature-based predictors. We selected 10 regions of interest (ROIs), including 196 vertices (Fig. 1C, see Materials and Methods for details) based on speech production and perception studies (Hertrich et al., 2020; Hickok & Poeppel, 2007; Indefrey & Levelt, 2004). These ROIs encompass the bilateral inferior frontal gyrus (IFG), precentral/postcentral gyri (Pre/PostCG), supramarginal gyrus/inferior parietal lobule (SMG/IPL), middle and inferior temporal gyri (MTG/ITG), and superior temporal and Heschl's gyri (STG/HG). We then selected an 8-second speaker-listener time-lag window for exploration to determine time lag in NS, ranging from -4s (the speaker's MEG responses precede those of the listeners by 4s) to 4s (listeners' responses precede those of the speaker by 4s), in a step of 10 ms.

The speaker's neural responses significantly predicted the listeners' responses across distributive region pairs (see Fig. 2A). We used a permutation test with 10,000 iterations to determine the significant threshold of the speaker-to-listener prediction (see Fig. 2B and Methods for details). We then identified 9226 speaker-listener regional pairs (157 from the speaker and 190 from the listeners) that showed significant interbrain prediction (see Fig. 2A). These region pairs were distributed across the 10 ROIs (Fig. 2C) with visually salient in the speaker's right frontotemporal-parietal regions and listeners' bilateral temporal regions (Fig. 2D). Across all region pairs, the speaker's RSTG best predicted the neural responses of listeners' bilateral STG (post hoc *t*-tests, corrected $p < 0.05$ [correction was applied for all following pairwise comparisons]; see Fig. 2D & 2E), followed by the speaker's right IFG and IPL.

We analyzed the NS for speaker and listener regions separately to better understand the effects of hemisphere and regional hierarchy on NS patterns. A significant hemisphere effect was found in the speaker regions ($F_{1, 210} = 300.1$, $p < 0.001$), with the right hemisphere exhibiting stronger NS than the left ($t_{21} = 19.3$, $p < 0.001$) (Fig. 2D & 2E, left panel). No significant hemisphere effect was observed for the listener regions ($F_{1, 210} = 2.43$, $p = 0.12$) (Fig. 2E, right panel). There was a significant main effect of ROI for both speaker ($F_{4, 210} = 28.8$, $p < 0.001$) and listener regions ($F_{4, 210} = 24.4$, $p < 0.001$), indicating that these ROIs have different contributions to NS (Fig. 2D-F).

We observed distinct NS patterns across ROIs between hemispheres for speaker regions, demonstrated by a significant interaction effect between hemisphere and ROI ($F_{4, 210} = 10.46$, $p < 0.001$). In the left hemisphere, temporal regions (STG/HG and MTG/ITG) predicted listeners better than frontal regions (IFG and Pre/PostCG) (post hoc $t$-tests, $p < 0.05$). Conversely, in the right hemisphere, temporal regions (STG/HG and MTG/ITG) outperformed prefrontal and parietal regions (Pre/PostCG and SMG/IPL) (post hoc $t$-tests, $p < 0.05$), except for IFG (post hoc $t$-tests, $p > 0.05$) (Fig. 2E, left panel). For listener regions, STG/HG displayed higher NS than the other four ROIs (post hoc $t$-tests, $p < 0.05$), and similar patterns were found across hemispheres ($F_{4, 210} = 0.15$, $p = 0.96$) (Fig. 2E, right panel).

These findings highlight that the strongest NS arises from channels between the speaker's right hemisphere, particularly in frontotemporal regions, and the listeners' bilateral auditory temporal cortices, emphasizing the importance of these interbrain neural channels for effective speaker-listener communication and representation sharing.

**Temporal dynamics of speaker-listener NS**

We examined the speaker-listener NS time-lag dynamics by analyzing TRF weights over time-lag windows (-4s to 4s). This analysis reflects the latency in predicting the listeners' neural responses based on the speaker's neural activity. Negative time lags indicate that the speaker's neural responses precede those of the listeners.

Most regions, except right Pre/PostCG, were significantly speaker-led ($t$-tests, $p < 0.05$), indicating that earlier neural responses from the speaker predicted later reactions from listeners. This speaker-led NS pattern was observed only in speaker-listener NS modeling; Zero time lag was found between listener pairs (see Fig. S5 for listener-listener NS patterns). For the speaker regions, we found significant time-lag windows of -1350 ms to 350 ms in the left hemisphere ($t$-tests, $p < 0.05$), with an average peak at -384 ms (Fig. 3A, middle-left and right panels). In the right hemisphere, the window ranged from -1250 ms to 750 ms, averaging -107 ms (Fig. 3A, middle-left and right panels). The left hemisphere's peak latencies occurred significantly earlier than the right's ($t_{21} = 6.34$, p < 0.001), suggesting it may be more involved in initiating speech production.

We also found a significant effect of ROI ($F_{4, 210} = 2.48$, $p < 0.05$) on peak latencies. In the left hemisphere, SMG/IPL (-276 ms) had less lag than other regions (ITG/MTG = -455 ms, STG/HG = -437 ms, IFG = -424 ms, Pre/PostCG = -326 ms) in predicting listeners' responses. In the right hemisphere, Pre/PostCG (59 ms) lagged behind other regions (ITG/MTG = -164 ms, STG/HG = 142 ms, IFG = -145 ms, SMG/IPL = -141 ms) (post hoc *t*-tests, $p < 0.05$) (Fig. 3A, left panels).

We also found that this hemisphere asymmetry in time-lag patterns varies between speakers and listeners (Fig. 3B & C). Speakers exhibited significant time-lag hemisphere asymmetry (left hemisphere significantly precedes right hemisphere) in IFG, Pre/PostCG, MTG/ITG, STG/HG (*t*-tests, $p < 0.05$) except for SMG/IPL (*t*-test, $p = 0.174$) (Fig. 3C, left panel), while listeners showed similar time lags between the two hemispheres (*t*-tests, $p > 0.05$; Fig. 3C, right panel). These time-lag differences indicate potential information flow across regions and hemispheres.

We further conducted principal component analysis (PCA) on TRF weights across time lag windows and region pairs (see Methods) to investigate the overall spatiotemporal patterns of speaker-listener NS. We identified 18 principal components (PCs) that explained 66.7% of the variance, outperforming permutation-defined chance (Fig. 3D). The first two PCs exhibited consistently positive scores (Fig. 3E) and distinct spatial patterns across region pairs (Fig. 3F). The first PC accounted for 34.0% of the variance, revealing a shared temporal NS pattern that began at -1240 ms (*t*-tests, $p < 0.05$) and peaked at -210 ms (Fig. 3E, left panel). Key region pairs contributing to this time-lag pattern included the speakers' bilateral temporal cortices and all listeners' regions, with a gradually increasing contribution from the speaker's frontal to the temporal areas (Fig. 3F, left panel).

The second PC explained 4.9% of the variance, highlighting differences between the speaker's two hemispheres, with significant positive scores from -1870 ms to -470 ms, peaking at -810 ms. This NS time-lag pattern indicated earlier left hemisphere activity compared to the right in predicting listeners' neural activity (Fig. 3E & F, right panel). These patterns suggest an information flow between speaker and listeners, from the left to the right hemisphere (PC2, earlier time lag: -810ms) and a hierarchical gradient structure within each hemisphere (PC1, later time lag: -210ms).

**Feature-driven speaker-listener neural synchronization**

By adding features to our content-based encoding modeling (Fig. 1B), we determined the extent to which different speech features and linguistic units contribute to the speaker-listener NS. This involved first estimating the speaker's neural signals induced by each of the six features of interest (i.e., acoustic, consonant, vowel, pitch height, pitch change, and tone category) to predict listeners' neural signals (see Fig. 4A). We estimated the unique variance explained (UVE [$R^2$]) of each feature to quantify the unique contribution of each feature to the speaker-to-listener neural prediction (i.e., NS) by controlling the variances induced by other features (see Materials and Methods for details).

Speakers' neural responses induced by the six features all uniquely and significantly predicted listeners' neural responses to different extents (Fig. 4B; permutation test, $p < 0.05$), indicating these features drive the emergence of inter-brain NS. Importantly, these features have distinct contributions to the NS (a significant main effect of feature; $F_{5,126} = 28.33$, $p < 0.001$). The tone category had the highest prediction and significantly outperformed other features, followed by vowel, pitch height, pitch change, and consonant, while acoustic features explained the least variances (post hoc *t*-tests, $p < 0.05$, Fig. 4B). The two segmental units (i.e., consonant and vowel) and the two pitch features (i.e., pitch height and change) did not significantly differ from each other in prediction (post hoc *t*-tests, $p > 0.05$, Fig. 4B).

These features not only contributed to the NS differently but also induced distinct speaker-listener synchronization patterns. We conducted pairwise feature classification analyses with a cross-validation procedure across all combinations of six features based on their NS patterns (10 speaker regions × 10 listener regions = 100 region pairs) (see Methods for details). All classification pairs, except between pitch height and pitch change (permutation tests, $p = 0.16$), had significant classification accuracy greater than chance (permutation tests, $p < 0.05$) (Fig. 4C, left panel), indicating different features induced distinct NS patterns. We visualized the feature-based interbrain NS pattern to the speaker and listener regions, respectively (Fig. 4D). We observed visually distinct synchronization patterns across features for the speaker regions, not only in the strength of NS but more importantly, in the spatial synchronization patterns distributed across regions. Pairwise feature classification analysis further confirmed the observation. All feature classification pairs based on the speaker's NS spatial patterns exhibited significant classification accuracies greater than chance (permutation tests, $p < 0.05$) (Fig. 4C, middle panel). In contrast, different features induced similar NS patterns across the listener regions (permutation tests, $p > 0.05$) (Fig. 4C, right panel). These features mainly differ in NS effect size, where more salient in the bilateral temporal cortices induced by tone category and vowel compared to others.

We further used PCA to identify the underlying components of the feature-based NS patterns. We found two significant PCs, explaining 87.6% of total variances (Fig. 4E). The first PC (80.6% variance explained) reflected the overall contribution to the NS (Fig. 4F, left panel), with the highest PC scores for tone category, followed by vowel, pitch height, pitch change, consonant, and acoustics. In contrast, PC 2 (7.0% variance explained) played an important role in weighting pitch cues followed by segmental units (Fig. 4F, right panel). The combination of the two PCs created a 2-dimensional NS space (Fig. 4G), which could significantly distinguish the six features in NS, placing the tone category on the bottom right, pitch height and change on the top middle, and low-level acoustics on the left corner of the space, while segmental units in between.

The two PCs are subserved by distinct interbrain NS patterns. PC 1 showed positive neural loadings across all speaker-listener region pairs (Fig. 4H, left panel), with slightly lower weighting on the speaker's left IFG and Pre/PostCG. PC 2 highlighted the greater loadings in the speaker's bilateral IFG and left Pre/PostCG (Fig. 4H, right panel). Combining the two PCs in the neural loading space (Fig. 4I) showed distinct contributions of different regions to the NS, where the left IFG and Pre/PostCG were distinct from other temporoparietal regions for the speaker-listener NS. These dissociations between speaker and listeners, as well as between left frontal and temporoparietal regions, demonstrate different mechanisms in achieving representation-sharing for distinct linguistic and speech features.

## Spatiotemporal patterns of feature-based neural synchronization

We examined the temporal dynamics of feature-based NS by analyzing the TRF weights for six feature estimations across time lags from -2s to 2s (Fig. 5A), covering significant windows indicated by the direct speaker-listener NS analysis. The TRF patterns of these features were like the direct TRF results shown in Figures 3A and 3B. To further explore the feature-driven spatiotemporal NS patterns, we conducted principal component analysis (PCA) on the six features. Fourteen principal components (PCs) accounted for 84.6% of the variance in feature-specific spatiotemporal dynamics (permutation test, $p < 0.05$) (Fig. 5B). Five PCs (PC 1, 2, 8, 11, and 14, explaining 54.2% variance) showed significantly different PC scores among features (one-way ANOVA, $p < 0.05$) (Fig. 5C) and exhibited significant positive neural loadings over time lags (*t*-tests, $p < 0.05$) (Fig. 5D), indicating these five PCs have distinct spatiotemporal NS patterns and substantially contributing to the overall NS.

The five PCs showed distinct weights between features and unique spatiotemporal NS patterns. PC 1 (30.1% variance explained) and PC 2 (19.4% variance explained) highlighted the contributions of the tone category on NS. For PC 1, scores for tone category were significantly higher than for acoustics and consonants (post hoc *t*-tests, $p < 0.05$) (Fig. 5C, first panel). For PC 2, tone category scores were significantly higher than those for the other five features (post hoc *t*-tests, $p < 0.05$) (Fig. 5C, second panel). The NS patterns of PC 1 and PC 2 began at -990 ms and -800 ms (*t*-tests, $p < 0.05$), peaking at -270 ms and -210 ms, respectively. The dominant region pairs for the two PCs involved the speaker's bilateral temporal cortices and right IFG, with listeners' left (PC 1) and right temporal cortices (PC 2) (Fig. 5E, first and second panels). These patterns can be seen as two latent components from the first PC of direct speaker-listener NS (Fig. 3E&F, left panels).

PC 8 (2.4% variance explained) demonstrated the contributions of pitch height and change on NS. In this PC, pitch height scored significantly higher than consonants, and all other features scored higher than the tone category (post hoc *t*-tests, $p < 0.05$) (Fig. 5C, third panel). This NS pattern started at -1170 ms (*t*-tests, $p < 0.05$) and peaked at -850 ms (Fig. 5D, third panel). The dominant region pairs were between the speaker's left regions and the listeners' left temporal regions, with a trend from the speaker's temporal to frontal regions (Fig. 5E, third panel). PC 8's spatiotemporal pattern coincided with the PC 2 from direct speaker-listener NS (Fig. 3E & F, right panels).

PCs 11 (1.3% variance explained) and 14 (1.0% variance explained) highlighted the consonant feature's contribution to NS. In PC 14, consonant's PC scores were significantly higher than other features (post hoc *t*-tests, $p < 0.05$) (Fig. 5C, fourth panel). For PC 11, the difference was not significant (post hoc *t*-tests, $p > 0.05$) despite having the highest mean score (Fig. 5C, fifth panel). Both PCs exhibit several peaks at

-550 ms, -230 ms, and 240 ms, with only the 240 ms peak being significantly greater than zero (*t*-tests, $p < 0.05$, FDR-corrected) (Fig. 5D, fourth and fifth panels). PC 11 was mainly influenced by the speaker's right parietal and temporal regions (Fig. 5E, fourth panel), while PC 14 involved the speaker's right temporal and left frontal regions (Fig. 5E).

Overall, distinct spatiotemporal NS patterns were identified for those features, subserving distinct neural communication channels. The tone category was primarily reflected between the bilateral temporal lobes and right frontal lobe of the speaker and the left (-270 ms) and right (-210 ms) temporal regions of the listeners. The NS pattern for pitch height and pitch change was mainly between the speaker's left hemisphere regions and the listeners' left temporal regions, peaking at -850 ms. The consonant NS pattern was primarily between the speaker's left frontal and right temporal regions and the listeners' bilateral temporal regions, with multiple peaks at -550 ms, -230 ms, and 240 ms.

**Speaker-listener neural synchronization predicted communication success**

We investigated whether NS between speakers and listeners could predict individual differences in communication success by analyzing comprehension accuracy following story listening. Using all speaker-listener region pairs, we constructed ridge regression predictive models to predict comprehension scores with a leave-one-subject-out cross-validation procedure (Fig. 6A). In this approach, each iteration used all subjects except one for training, and the held-out subject for testing, systematically rotating through all participants. One subject was excluded from analysis as an outlier due to comprehension scores significantly below others.

The feature-driven NS patterns demonstrated robust predictive power for listeners' story comprehension scores (Fig. 6B). The model achieved high accuracy with a mean squared error of 0.00094 and an $R^2$ value of 0.347 ($p = 0.003$), substantially exceeding the permutation-based significance threshold ($p = 0.05$, $R^2 = 0.079$). As illustrated in the scatter plot (Fig. 6B, middle panel), predicted comprehension scores closely matched actual performance across subjects, while the histogram (Fig. 6B, right panel) confirms that this prediction accuracy exceeded chance levels. In addition, different feature-driven NS varied differently in predicting comprehension scores (see Table S1 in Supplementary).

Different linguistic features also exhibited distinctive NS patterns that contributed to predicting communication success (Fig. 6C). These NS maps revealed feature-specific predictive contribution patterns. Tonal features ("tone") and pitch change and height cues ("pc" and "ph") showed predominantly positive coefficients (red/yellow connections) between the right temporal and frontal regions of speakers and the right temporal regions of listeners. Consonant ("con") and vowel ("vow") features displayed strong positive coefficients, primarily between the right temporal regions of speakers and the right temporal regions of listeners. Acoustic features ("ac") exhibited a more complex pattern with positive coefficients between bilateral frontal and parietal regions of speakers and right temporal regions of listeners, alongside negative coefficients between speakers' right temporal regions and listeners' left hemisphere regions. Phonological features ("ph") showed distributed synchronization patterns across multiple brain regions.

These findings demonstrate that NS between speakers and listeners, particularly involving right-hemisphere frontotemporal regions, serves as a robust neuromarker for successful communication, with language-specific tone-related features engaging distinct underlying patterns of inter-brain synchronization.

## Discussion

This study examined how language-specific and language-general features shape speaker-listener neural synchronization (NS) in Mandarin Chinese. By combining a pseudo-hyperscanning magnetoencephalography (MEG) paradigm and feature-based interbrain synchronization modeling, we characterized the spatiotemporal dynamics of NS and assessed the unique contributions of six linguistic and speech features, including acoustics, consonants, vowels, pitch height, pitch change, and tone category. Tone category and pitch features are predominant in driving NS, consistent with the critical function of pitch in Mandarin lexical processing. These results support the view that effective verbal communication relies on shared neural representations of linguistic content, with language-specific features exerting substantial effects on speaker-listener alignment in tone languages.

Our findings revealed strong speaker-listener NS across various cortical regions with different time lags. The left frontotemporal regions were crucial for initiating NS at longer delays (around -800 ms). In comparison, bilateral temporoparietal regions significantly facilitated NS at shorter delays (about -200 ms) between speaker and listener responses. Notably, tone-language-specific features, particularly tone category and associated speech cues (like pitch height and change), were vital in driving NS. These tonal features showed unique spatiotemporal NS patterns compared to language-general features such as syllables and segmental units. These findings suggest that users of tone languages, like Mandarin, have specialized neural mechanisms for sharing tonal information during communication. Importantly, these language-specific NS patterns were significant predictors of listeners' story comprehension scores, a key measure of successful communication, surpassing the predictive power of NS from acoustic features.

**The pivotal role of lexical tones in driving NS**

Our findings support and significantly extend the representation-sharing hypothesis of communication, which posits that effective verbal exchange depends on shared neural representations of speech content between speakers and listeners (Glanz et al., 2018; Hasson et al., 2012; Jiang et al., 2021). Recent research into neural coupling has provided preliminary evidence that such sharing unfolds across multiple processing levels, from low-level acoustic features to high-level semantics (Li et al., 2024). Here, we confirmed that the speaker's neural responses significantly predicted those of the listeners, revealing systematic synchrony patterns spread across cortical regions and time intervals. We also demonstrated that different speech cues and linguistic features contribute uniquely to these predictions (i.e., NS). This feature-specific interbrain alignment underscores a sophisticated temporal coordination mechanism operating at multiple linguistic tiers.

By focusing on specific speech and linguistic features, we found that tone-related features, specifically tone categories and pitch cues, exert a stronger influence on NS than non-tonal cues, such as consonants, vowels, or the acoustic envelope. This aligns with earlier work pointing to the preferential processing of tonal characteristics in both production and perception networks for tone language speakers (Berthelsen et al., 2020; Chien et al., 2020; Lu et al., 2023; Sereno & Lee, 2015). Consequently, our results highlight the critical role of pitch and tonal information in communication with tone languages, where variations in pitch contour carry lexical significance (Gandour et al., 2000; Y. Li et al., 2021). Moreover, the apparent prioritization of tonal features is likely supported by specialized fronto-temporoparietal processing networks that are specially developed in tone language speakers (Chien et al., 2020; Feng, Gan, Llanos, et al., 2021; Feng et al., 2018; Feng et al., 2019). Indeed, these networks exhibit enhanced neural representations of tone categories even after short-term category training, suggesting an experience-dependent optimization of the neural architecture for pitch-relevant information processing (Feng, Li, et al., 2021; Feng et al., 2019).

The predominance of tone category and related pitch cues in driving NS (accounting for 87.6% of NS variance) indicates that speakers and listeners are particularly attuned to pitch variations in tone languages, thereby facilitating synchronized neural activity. Because pitch changes fundamentally alter syllabic meaning, accurate pitch production and perception are vital for successful communication. Hence, the neural system may prioritize tone-specific information, leading to robust interbrain synchronization during lexical exchange. These findings underscore that language-specific features play a key role in shaping the brain's capacity to align neural activity beyond the influence of more general auditory and linguistic features. This highlights human neural architecture's adaptability to the unique demands of diverse languages.

The substantial contribution of tonal features to NS emerged across multiple cortical regions, with an especially pronounced link between the speaker's fronto-temporoparietal regions (mainly in the right hemisphere) and the listener's bilateral auditory-motor areas (Fig. 4D). These regions play important roles in planning and executing speech processes, including the precise control of pitch movements necessary for tone articulation (Chang et al., 2013; Chien et al., 2020; Dichter et al., 2018) and the multi-dimensional representations of pitch and tone categories for perception (Feng, Gan, Llanos, et al., 2021; Feng et al., 2018; Feng et al., 2019). The involvement of the right hemisphere in encoding tone-related features aligns with its roles in processing linguistic and phonological information (Alexandrou et al., 2017; Ozker et al., 2022; Tourville & Guenther, 2011). Bilateral engagement, in turn, may reflect shared representations of tonal features, requiring both precise pitch control for production and fine-grained category perception (Binder et al., 2000; Nan & Friederici, 2013; Tang et al., 2017).

These spatial NS patterns reveal a refined interplay between production and perception networks, potentially optimized for tonal language processing. The pronounced right-hemispheric activation observed in speakers, coupled with bilateral engagement in listeners, suggests a specialized neural architecture attuned to the unique demands of tonal communication. Our feature-driven analyses also revealed how cross-frequency interactions could underlie the distinct spatiotemporal patterns of NS for pitch-based features compared to consonantal or acoustic cues. The broader literature on speech-brain entrainment emphasizes that gamma-band synchronization often encodes short-lived phonetic detail (e.g., formant transitions or voice-onset time), while theta/delta oscillations manage syllabic grouping or prosodic/intonational structure. In tone languages, this multi-tier system may become even more specialized. A pitch contour spans hundreds of milliseconds (roughly a syllable), suggesting that theta-band phase-locking may carry pitch contour information, while faster oscillations encode moment-to-moment spectral details (Liang & Du, 2018). Together, these nested oscillations foster a hierarchical neural parsing of continuous speech, indicating that tone categories occupy an intermediary timescale, “slower” than a single phoneme but “faster” than full intonation contours. Thus, the bilateral but right-dominant frontotemporal networks observed may reflect a cross-frequency bridging of pitch, segmental articulation, and higher-level lexical functions. Future studies should examine how these patterns differ in speakers of non-tonal languages and how they evolve throughout tone language acquisition.

### Information flow of speaker-listener neural synchronization

We extended our analysis of NS to pinpoint the precise time-lag relationships between the speaker’s and listeners’ shared neural responses. Consistent with prior work (Chang et al., 2024; Liu et al., 2020; Liu et al., 2017; Stephens et al., 2010), our data show that the speaker's neural activities mainly precede the listener's. Importantly, these time lags varied across region pairs and speech features, revealing directional information flow both within and across brains.

For the tone category, we observed a robust temporal lead (~ -250 ms) emerging from the speaker's bilateral temporal regions and right inferior frontal gyrus (IFG) to the listeners' bilateral temporal regions. This pattern likely reflects the efference copy mechanism, in which the brain generates an internal representation copy of the speech commands and sends it to auditory regions as a prediction of the expected sensory consequences of the impending production (Hickok et al., 2011; Li et al., 2020; Ozker et al., 2022). These production representations convey information through multi-dimensional tone category encoding for listeners' efficient perception (Feng, Gan, Llanos, et al., 2021; Feng et al., 2018; Feng et al., 2019). In contrast, pitch cues showed an earlier lead (~ -850 ms) from the speaker's left IFG to the listeners' left superior temporal gyrus (STG). These earlier synchronized activations are consistent with the left-hemisphere role in linguistic encoding and articulatory planning (Hickok & Poeppel, 2007; Indefrey & Levelt, 2004), which requires accumulating pitch heights across times to signal contour patterns and abstract tone categories. Together, these distinct spatiotemporal patterns for tone category and pitch cues align with the direct speaker-listener NS results (Fig. 3E & 3F), underscoring how language-specific pitch-based features can shape both when and where communicative alignment occurs.

In contrast to segmental features (e.g., consonants/vowels), tones require continuous pitch tracking over the syllable to establish lexical identity. This may explain the robust speaker-led synchronization we observed at longer time lags (∼ -850 ms for pitch cues and ∼ -250 ms for discrete tone categories). These earlier speaker-to-listener leads potentially reflect production-based predictive coding mechanisms in which the speaker's motor commands for pitch are internally modeled and then propagated forward (Hickok et al., 2011; Ozker et al., 2022). When listeners' cortical rhythms align with these tone-centric predictive signals, mutual neural entrainment arises, facilitating more precise and rapid mapping of lexical information.

Consonant-related NS, by contrast, emphasized the speaker's left IFG and right temporal regions and listeners' bilateral temporal regions; here, listeners' neural activity actually preceded the speaker's (~ 240ms), possibly reflecting predictive mechanisms on the listener's part (Dai et al., 2018; J. Li et al., 2023). This anticipatory aspect adds to the larger picture of feature-dependent synchronization, demonstrating how different linguistic cues may engender distinct spatiotemporal dynamics and the directionality of speaker-listener information exchange.

**Implications for the neural mechanisms of verbal communication**

Our study offers novel insights into how various speech and linguistic features contribute to speaker-listener NS. The differential impact of speech features on NS, particularly the dominance of tonal cues, underscores a hierarchical organization of neural processing across production and perception (Brodbeck et al., 2022; Evans & Davis, 2015; Fairs et al., 2021; Jiang et al., 2021; Sheng et al., 2019). In line with our earlier feature-driven results, the prominent role of tone-related information emphasizes the need to incorporate language-specific properties into models of communication processing. This feature-dependent synchronization suggests that neural coupling mechanisms are finely tuned to linguistically salient inputs, operating with high temporal precision (Z. Li et al., 2023; Pastore et al., 2022) to enable seamless speaker-listener alignment.

These findings expand our theoretical understanding of neural coupling during communication in several ways. First, they reinforce our hypothesis that NS unfolds via feature-specific processing streams, as demonstrated by the spatial and time-lag pattern analyses. Second, the predominance of tonal features aligns with frameworks proposing that neural coupling prioritizes linguistically critical cues (Jiang et al., 2021; Li et al., 2024). The earlier activation of the left hemisphere in speakers, coupled with stronger right-hemisphere NS, points to a dynamic interplay between hemispheres that is tailored to the unique demands of tonal speech communication. This hemispheric specialization appears to reflect an optimized processing strategy for integrating multiple linguistic features simultaneously (Alexandrou et al., 2017; Floegel et al., 2020; Schulz, 1997), as further evidenced by our time-lag findings in tone category and pitch cues.

In keeping with these observations, the observed patterns of NS point to a sophisticated neural architecture underlying content-based representation sharing.

Recent neuroimaging studies suggest that this architecture leverages both sequential and parallel processing routes optimized for distinct linguistic features (Fairs et al., 2021; Gao et al., 2024). Our demonstration of the right-to-both-hemispheric NS patterns in tone and pitch processes paired with left-hemispheric involvement in other speech components offers a complementary model that efficiently handles the complexity of spoken language (Kelsen et al., 2022). Such bilateral coordination is presumably facilitated by rapid interhemispheric communication, allowing real-time integration of multiple speech features, as highlighted in our feature-based TRF analyses. Building on these mechanistic insights, our findings have practical implications for understanding communication disorders. The feature-specific nature of NS raises the possibility that communication deficits may stem from disruptions in particular processing streams, rather than a general inability to synchronize. This view opens new clinical and therapeutic pathways centered on targeting the specific linguistic features implicated in a given speech disorder (Kent & Kim, 2003; Pennington & Bishop, 2009). For example, individuals with tone-perception impairments may benefit from interventions emphasizing pitch awareness and production training, drawing on our demonstration that tonal features can be pivotal for speaker-listener alignment.

## Conclusion

In summary, these findings demonstrate that effective verbal communication depends on shared neural representations of speech content, with language-specific tonal features playing a crucial role in interbrain neural alignment in tone languages. It also deepens our understanding of the neural mechanisms behind speech communication by clarifying the spatiotemporal dynamics of neural synchrony and how language-specific traits influence the interaction and synchronization of interlocutors' speech production and perception systems. These findings open new pathways for investigating how language diversity impacts neural coupling during communication and could guide the development of more effective communication strategies in educational and clinical contexts.

### Limitations and future directions

Despite these contributions, several limitations should be acknowledged. First, our pseudo-hyperscanning design, where the speaker and listeners were recorded separately, cannot capture the interactive dynamics characteristic of real-time, face-to-face communication. Future research employing simultaneous hyperscanning techniques, such as dual-MEG during natural dialogue, could offer deeper insights into the neural mechanisms supporting live speaker-listener interactions. Second, our study focused solely on Mandarin Chinese, leaving the question of whether different NS patterns might be observed in non-tonal languages unresolved. Comparative studies across diverse languages would clarify the universality and specificity of the neural processes underlying communication. Lastly, although we concentrated on sublexical features, higher-level linguistic components, including syntax and semantics, also significantly influence speech perception and production and their synchronizations in the brain. Future investigations should examine how these higher-order elements interact with sublexical cues to modulate NS and foster shared representations between speakers and listeners.

## Materials and Methods

### Participants

A total of 28 participants participated in this study, consisting of six speakers (3 females; mean age: 23.3 years; age range: 22-25 years; only the audio recordings and MEG data from a 23-year-old female speaker were used for the listening) and 22 listeners (12 females; mean age: 25.6 years; age range: 21-37 years). All participants were native Mandarin speakers with self-reported normal hearing and no history of neurological or psychiatric disorders. Participants provided written informed consent prior to the study and were compensated for their participation. The study was approved by the Joint Chinese University of Hong Kong (CUHK) - New Territories East Cluster (NTEC) Clinical Research Ethics Committee (CREC) and Research Ethics Committee at National Taiwan University.

### Experimental Procedure

The experiment consisted of two main parts: story-telling by the speaker and story-listening by the listeners (Fig. 1A). To create a pseudo-hyperscanning paradigm, the stories narrated by a speaker were recorded and subsequently played back to the listeners during MEG scanning. This sequential design ensured consistent auditory stimuli across listeners and accommodated the limitations of MEG equipment regarding simultaneous scanning.

During the story-telling session, the speaker was instructed to read aloud eight short stories presented on a screen and then spontaneously retell each story from memory. The stories were adapted from children's narratives and adjusted for length and complexity. During story presentation, text appeared on the screen sentence by sentence, with each sentence containing 6 to 37 words, presented for 2.4 to 13.9 seconds to maintain natural reading pace.

After reading each story, the speaker had a few minutes to prepare before retelling the story aloud, without reading from the text. The retelling phase was recorded using a microphone at a sampling rate of 48 kHz and 16-bit resolution. Only the recordings and MEG signals from the retelling phase were used in the subsequent story-listening sessions and analyses.

In the story-listening session, each listener underwent eight trials, corresponding to the eight stories told by the speaker. The recorded narratives were played back through air-tube earphones at a comfortable volume (60-70 dB SPL) to prevent potential electromagnetic interference with the MEG signals.

Listeners were instructed to listen attentively to each story when a fixation cross was displayed on the screen to minimize eye movements and blinks. After each story, listeners answered three yes/no comprehension questions presented visually on the screen and responded using a button press to ensure engagement and understanding. At the end of all stories, listeners completed an additional 56 comprehension questions (7 per story) to test their story understanding and recall.

**Data Acquisition and Preprocessing**

*Neuroimaging recordings*

MEG data were collected using a whole-head 306-channel system (Elekta Neuromag TRIUX), which includes 204 planar gradiometers and 102 magnetometers, with a sampling rate of 1000 Hz. The MEG system was in a soundproof and magnetically shielded room at the Imaging Center for Integrated Body, Mind, and Culture Research, National Taiwan University. Structural MRI scans were collected for each participant using a 3T MRI scanner (Siemens Trio) employing a T1-weighted MPRAGE sequence (repetition time = 2.53 s; echo time = 3.37 ms; field of view = 256 mm; voxel size = 1 $mm^3$). These scans were used to develop individual head models for MEG source localization.

*Preprocessing*

MEG data were preprocessed using MaxFilter software (Elekta Oy, Finland) for signal-space separation, effectively removing external noise and compensating for head movements. The data underwent band-pass filtering between 0.3 and 45 Hz using a one-pass zero-phase finite impulse response (FIR) filter (order 6038, kaiser-windowed sinc FIR, with a sampling rate of 1000 Hz) to eliminate slow drifts and high-frequency noise. After filtering, the speaker's MEG signals (Fig. S1C & D, top panels) were slightly noisier than the listeners' (Fig. S6A). However, this signal difference was not contributed by the speaker's vocalization actions (Fig. S1A & B). The identified noise likely stemmed from involuntary body and eye movements. Among the six speakers analyzed, one was chosen for the story-listening experiment based on the quality of the MEG data and the fluency of the narration.

Independent component analysis (ICA) was employed to identify and eliminate artifacts commonly introduced by eye movements, blinks, muscle movements and cardiac activity (Vigário et al., 2000). Each independent component (IC) was closely examined using multiple criteria, including its topography (peak weights over specific sensors) and its time-course properties (e.g., square waves or pulses for blinks, ~1 Hz pulsation for cardiac activity). Eye-movement and blink components often showed maximal weights located at the bilateral frontal sensors, with spiking or pulsed waveforms corresponding to saccades or blinks. Cardiac components typically displayed peak activation near the bilateral occipital sensors (close to neck/vascular areas) and a characteristic 1 Hz rhythm matching the individual's heart rate. Muscle movements resulted in peaks at bilateral temporal regions and showed fluctuation synchronized to sound envelope (Pearson's r value of ~0.1). Components identified as artifacts (ranging from 1 to 5 ICs) were removed from the dataset by zeroing those specific IC activations. The remaining ICs, representing predominantly neural activity of interest, were then projected back into sensor space, providing a cleaner dataset for subsequent analyses. After that, the correlation between the MEG signal and the sound envelope decreased, implying that the muscle noise associated with vocal was largely removed (Fig. S1C &D). This ICA-based procedure substantially attenuated non-neural contributions, thereby improving the signal-to-noise ratio.

To align the MEG signals with the speech recordings, we used an MEG channel to record the audio stimuli presented during the experiments. The speech stimuli were downsampled to 1000 Hz (the MEG sampling rate) and temporally aligned with the MEG signals by maximizing the correlation between the spectrograms of the original stimuli and the recordings obtained during the MEG experiments. The MEG data were then segmented to align with the duration of each story and downsampled to 100 Hz for computational efficiency.

*Source Reconstruction*

MEG source reconstruction was conducted using MNE 1.5.1 (Gramfort et al., 2014) under Python 3.11.6. Individual head models were generated from T1-weighted MRI data using a single-layer boundary element model (BEM) (Fuchs et al., 2002) by FreeSurfer v7.4.0. Co-registration of MEG sensors with MRI data was accomplished using three anatomical fiducials: the nasion and the bilateral preauricular points. To implement the reconstruction, we performed the forward solution to create a model that describes the magnetic fields at the measurement sensors produced by dipole sources on the cortex, followed by the inverse solution to estimate the activation at the source level. The forward solution was computed using a source space containing approximately 8192 vertices per hemisphere (spacing = 5 mm). The inverse solution was calculated using dynamic statistical parametric mapping (dSPM) (Dale et al., 2000) with loose orientation constraints (loose = 0.2) and depth weighting (depth = 0.8), assuming a signal-to-noise ratio (SNR) of 1. Individual source estimates were then morphed to the FreeSurfer "fsaverage" brain template for group analyses. Based on the parcellation derived from functional connectivity (Schaefer et al., 2018), source signals were organized into 1000 whole-cortical parcels. For further analysis, we focused on 196 parcels from brain regions associated with speech production and perception (Fig. 1C), including the inferior frontal gyrus (IFG; number of parcels: left 10, right 13), precentral and postcentral gyri (Pre/PostCG; number of parcels: left 15, right 14), supramarginal gyrus/inferior parietal lobule (SMG/IPL; number of parcels: left 28, right 19), middle and inferior temporal gyri (MTG/ITG; number of parcels: left 18, right 19), and superior temporal gyrus/Heschl's gyrus (STG/HG; number of parcels: left 30, right 30).

## Speech Feature Extraction

*Transcription and linguistic feature labeling*

A native Mandarin speaker with transcription experience was recruited to perform initial transcriptions of eight audio recordings of stories. To enhance accuracy, two additional native Mandarin speakers acted as independent verifiers, cross-checking the transcriptions against the original recordings. The primary transcriber utilized speech-to-text software to generate initial transcripts, capturing all verbal content. The verification process involved the verifiers listening to the audio recordings and comparing them with the transcripts, noting any discrepancies for further examination. A collaborative meeting was held to discuss and resolve these discrepancies, ensuring that the final transcripts accurately reflected the audio content. Key linguistic features of interest were identified for annotation, including the onsets and durations of consonants, vowels, tones, syllables, characters, words, and sentences. All participants underwent training to standardize feature labeling, which was conducted using Praat (Boersma, 2011). to maintain precision and consistency. Inter-rater reliability was assessed through direct comparisons between raters. Items with timing differences greater than 100 milliseconds were relabeled by the annotators to ensure a high level of agreement. Follow-up meetings addressed any discrepancies in annotations, facilitating consensus building. A comprehensive final review confirmed the accuracy and completeness of the transcriptions and annotations, with all decisions documented for transparency.

*Acoustic features*

Two acoustic features (i.e., envelope and acoustic onset) were extracted from speech recordings using an eight-band gammatone filter bank, which spans frequencies from 20 Hz to 5000 Hz on an equivalent rectangular bandwidth scale. The envelope for each frequency band was extracted and adjusted using an exponent of 0.6 to better align with human loudness perception (Brodbeck et al., 2023). The overall acoustic envelope was derived by averaging the envelopes across all bands. Additionally, the acoustic onset was computed by taking the first derivative of the acoustic envelope and rectifying the resulting signal.

*Phonetic features*

The phonetic features analyzed in this study included the onsets of consonants and vowels, and durations of syllables. From the linguistic annotations, binary time-series predictors were constructed for the encoding modeling. The value was set to '1' at the onset of a consonant or vowel, and '0' at other timepoints. At each time point in the recordings, the presence or absence of a syllable was indicated by a value of '1' or '0', respectively. The reason we described syllables in terms of durations is to facilitate comparisons with the tone-related features below.

*Pitch features*

The fundamental frequency (F0) of the speech recordings was estimated using the autocorrelation method implemented in the Praat software package (Boersma, 2011). Prior to analysis, the recordings were preprocessed to minimize background noise and ensure optimal pitch detection. Manual corrections were performed to address common pitch tracking errors such as octave jumps and halving, enhancing the accuracy of the F0 measurements.

Pitch height was quantified by calculating the natural logarithm of the F0 values (log-F0), providing a perceptually linear representation of pitch consistent with human auditory perception. This transformation is crucial for accurately modeling how pitch height variations are perceived in spoken language.

To capture the dynamic aspects of pitch, the pitch change was calculated as the first temporal derivative of the log-F0 values (Y. Li et al., 2021). This measure reflects the rate of pitch change over time, highlighting pitch contour patterns essential in tone languages like Mandarin.

Both pitch height and pitch change values were discretized into ten equally spaced bins, covering the full range of observed values in the dataset. At each time point, these discretized values were represented as 10-dimensional binary vectors, where each dimension corresponded to a specific bin, and a value of '1' indicated that the pitch feature fell within that bin and '0' indicated that it did not fall in that bin. This discretization facilitated the incorporation of continuous pitch features into statistical and computational models.

*Tone category features*

Mandarin Chinese is characterized by its use of lexical tones, where variations in pitch contour signal tone categories and convey differences in word meaning. The tone category features labeled were based on the five Mandarin tones: high-level (Tone 1), rising (Tone 2), dipping (Tone 3), falling (Tone 4), and neutral (Tone 5).

Time intervals corresponding to each tone category were identified using the previously annotated syllable boundaries. For each tone, binary predictors were created wherein a value of '1' indicated the presence of that tone at a given time point, and a '0' indicated its absence.

**Speaker-Listener Neural Synchronization Modeling**

To quantify neural synchronization between the speaker and listeners during natural speech processing, we employed a multivariate temporal response function (mTRF) modeling approach using the mTRF Toolbox (Crosse et al., 2016). The mTRF method models the linear relationship between the neural signals of the speaker and listeners across various time lags, capturing the shared temporal dynamics inherent in speech production and perception processes.

*Direct Speaker-Listener Neural Prediction*

The direct neural synchronization between the speaker and listeners was analyzed using the TRF modeling framework (Fig. 1A). here, we used the listener's neural responses to predict the speaker's neural signals via a linear convolution, assuming that the MEG responses of the listener are temporally related to those of the speaker. The TRF model is defined as:

$$r(t,n) = \sum_{\tau} \omega(\tau,n)s(t-\tau) + \varepsilon(t,n)$$

where $r(t,n)$ is the neural responses of the listener at sensor $n$ and time $t$, $s(t-\tau)$ is the speaker's neural responses at time $t-\tau$, $\omega(\tau,n)$ is the TRF weight at timelag $\tau$ for sensor $n$, and $\varepsilon(t,n)$ is the residual error.

The TRF weights $\omega$ were estimated using ridge regression, which minimizes the mean squared error between the actual and predicted listener neural signals while applying a regularization term to prevent overfitting. A relatively broad range of time lags from -4000 ms to 4000 ms (sampled at 100 Hz) was considered exploratory to accommodate neural processing delays or predictions associated with speech perception. The regularization parameter (lambda) was set to $10^3$ to control the smoothness and avoid overfitting of the TRF weights $\omega$.

Model performance was evaluated by computing the Pearson correlation coefficient between the predicted and actual neural signals of the listener for each source location, yielding an encoding performance metric *r*. To assess the generalizability of the model across different speech content, leave-one-story-out cross-validation was performed, where the model was trained on all but one story and tested on the held-out story in an iterative manner for each subject.

For each listener, the model was individually trained for each of the speaker's MEG source locations to predict all 196 MEG locations of the listener. This resulted in an encoding performance matrix of dimensions 196 × 1 for the correlation coefficients and TRF weights of dimensions 196 × 801, corresponding to 801 timelags at a 100 Hz sampling rate over the -4000 ms to 4000 ms window.

*Feature-driven Speaker-to-Listener Neural Prediction*

To examine how specific speech features contribute to neural synchronization between the speaker and listeners, we employed a two-stage TRF modeling approach (Fig. 1B). In the first stage, we estimated the speaker's neural encoding of each speech feature by training TRF models with individual speech features as inputs and the speaker's MEG signals as outputs. The speech features included phonetic elements (e.g., consonants and vowels), pitch features (e.g., pitch height, pitch change), and tone categories specific to Mandarin Chinese.

The resulting feature-specific neural estimations represent the speaker's neural responses associated with the encoding of each feature. In the second stage, these estimated neural responses were used as inputs to predict the listeners' neural signals using TRF models, analogous to the direct speaker-listener prediction. To focus on the significant time-lag window observed in direct neural coupling, the time lags were restricted to -2000 ms to 2000 ms.

To quantify the unique contribution of each feature to neural synchronization (Tezcan et al., 2023), we first established a full model by training a TRF with all feature-specific neural estimations as inputs. Subsequently, we trained reduced models by systematically excluding one feature at a time. The unique variance explained (UVE) by each feature was calculated by subtracting the variance explained by the reduced model (without the feature) from that of the full model:

$$UVE = R^2_{full\ model} - R^2_{reduced\ model}$$

$$R^2(y, \hat{y}) = 1 - \frac{\sum_t (y_t - \hat{y}_t)^2}{\sum_t (y_t - \bar{y})^2}$$

where $\hat{y}_t$ is the prediction of signal at the timepoint $t$ and $y_t$ is the corresponding true value. $\bar{y}$ represents the mean of $y_t$ across all timepoints.

This approach isolates each feature's specific contribution to the neural coupling between the speaker and listeners. The temporal dynamics and spatial distributions associated with each feature were examined by analyzing the TRF weights from the full model.

**Statistical Analysis**

Statistical significance of the neural encoding and the unique contributions of speech features were assessed using permutation testing. The encoding performance metrics were compared against a null distribution generated by randomly shuffling the time segments of the input signals (2 to 102 seconds in steps of 0.1 sec), effectively disrupting the temporal relationship between speaker and listener signals. This process was repeated 1,000 times to establish a robust null distribution for significance testing.

To avoid multiple comparison problems, we used the maximum of 196*196 values obtained by each permutation to establish the null distribution (Maris & Oostenveld, 2007). The threshold for significance was set as the 95th percentile of the null distribution ($p < 0.05$). To include the brain sources involved in speaker-listener neural synchronization as completely as possible, we recruited all speaker/listener sources with significant synchronization to at least one listener/speaker source(s) for further analysis.

To correct for multiple comparisons across channels, regions and features, we applied the false discovery rate (FDR) correction (Benjamini & Hochberg, 1995). For other post hoc analyses, more stringent Bonferroni correction was used where appropriate.

**Classification of spatiotemporal pattern of NS**

To perform pairwise classification of features, the linear discriminant analysis (LDA) was conducted on UVE of all region pairs and speaker/listener regions. The UVE of speaker/listener regions were calculated by averaging across listener/speaker regions.

A leave-one-subject-out validation was utilized to avoid overfitting. The null distribution of accuracies was estimated by performing the same LDA procedure after randomizing the sample labels of features 1,000 times.

**Principal Component Analysis**

Principal component analysis (PCA) was performed on the TRF weights and unique variance explained to identify underlying patterns in the data. The data was arranged into a PCA data matrix of observation by variable. The observation dimension of TRF weights of direct NS was subject * timepoint (22 * 781), while for UVE and TRF weights of feature-driven NS, it was subject * feature (22 * 6) to investigate the spatial-temporal patterns of each feature. The variable dimension was speaker-listener regions pair (10 * 10 = 100) for TRF weights of direct NS and UVE of feature-driven NS and region pair * timepoint (100 * 381) for TRF weights of feature-driven NS. The number of significant components was first determined using permutation tests and the broken-stick method (Jackson, 1993). In each permutation, each column of the data matrix was scrambled individually. Bootstrap resampling (iterations = 10,000) was used to assess the significance of the loadings for each significant component. The components with significant loadings were recruited for further analysis.

**Predictive Modeling of Communication Success**

A predictive model was developed to explore the relationship between neural synchronization (NS) and the behavioral performance of listeners, specifically their accuracy in answering comprehension questions after listening to stories. The predictor variables in the model included feature-driven NS (feature*region pair=6*100). The feature-driven NS comprised the UVEs relevant to each linguistic feature (i.e., acoustic, phoneme, syllable, prosodic cues, vowel, and tone) used to ascertain how specific synchronization related to these features contributes to communication success.

A leave-one-subject-out cross-validation procedure was employed to ensure the model's robustness and generalizability (Fig. 6A). This approach involved using data from all but one subject pair to train the model, while the excluded subject's data served as an independent test set. This process was repeated until each speaker-listener subject pair had been tested once. The model was implemented in Python 3.11.6 using scikit-learn 1.5.0, incorporating a ridge penalty term to regularize the linear regression coefficients and avoid overfitting. The regularization parameter (lambda) was optimized over a logarithmic scale ranging from $10^{-20}$ to $10^{20}$ and the optimal lambda value is 2.8e-5 (Fig. 6B, left panel). This strategy ensured a selection of the optimal balance between bias and variance. One outlier subject was excluded from the prediction analysis due to abnormally low accuracy (72.5%), which could have otherwise distorted the regression coefficients and compromised the reliability of the model.

The performance of the model was assessed using two key metrics: Mean Squared Error (MSE) and R-squared ($R^2$). MSE measures the squared Euclidean distance between the predicted and actual behavioral accuracies. The R-squared indicates the proportion of variance in behavioral accuracy that the model explains. Null distributions were generated through permutation testing to assess the statistical significance of these measures. This process involved randomly shuffling behavioral accuracies across subjects and refitting the model (10,000 iterations) to generate the null. The observed $R^2$ or MSE value was then compared to the 95th percentile of the null distribution, resulting in a calculated p-value.

## Acknowledgments

## Funding

The work described in this paper was supported by grants from the Research Grants Council of the Hong Kong Special Administrative Region, China (Project No.: 14614221, 14612923, 14621424, and C4001-23Y to Gangyi Feng) and the National Natural Science Foundation of China (Project No. 32322090 to Gangyi Feng).

## Data availability statement

The raw data supporting the conclusions of this article will be made available by the authors.

## Figures and Tables

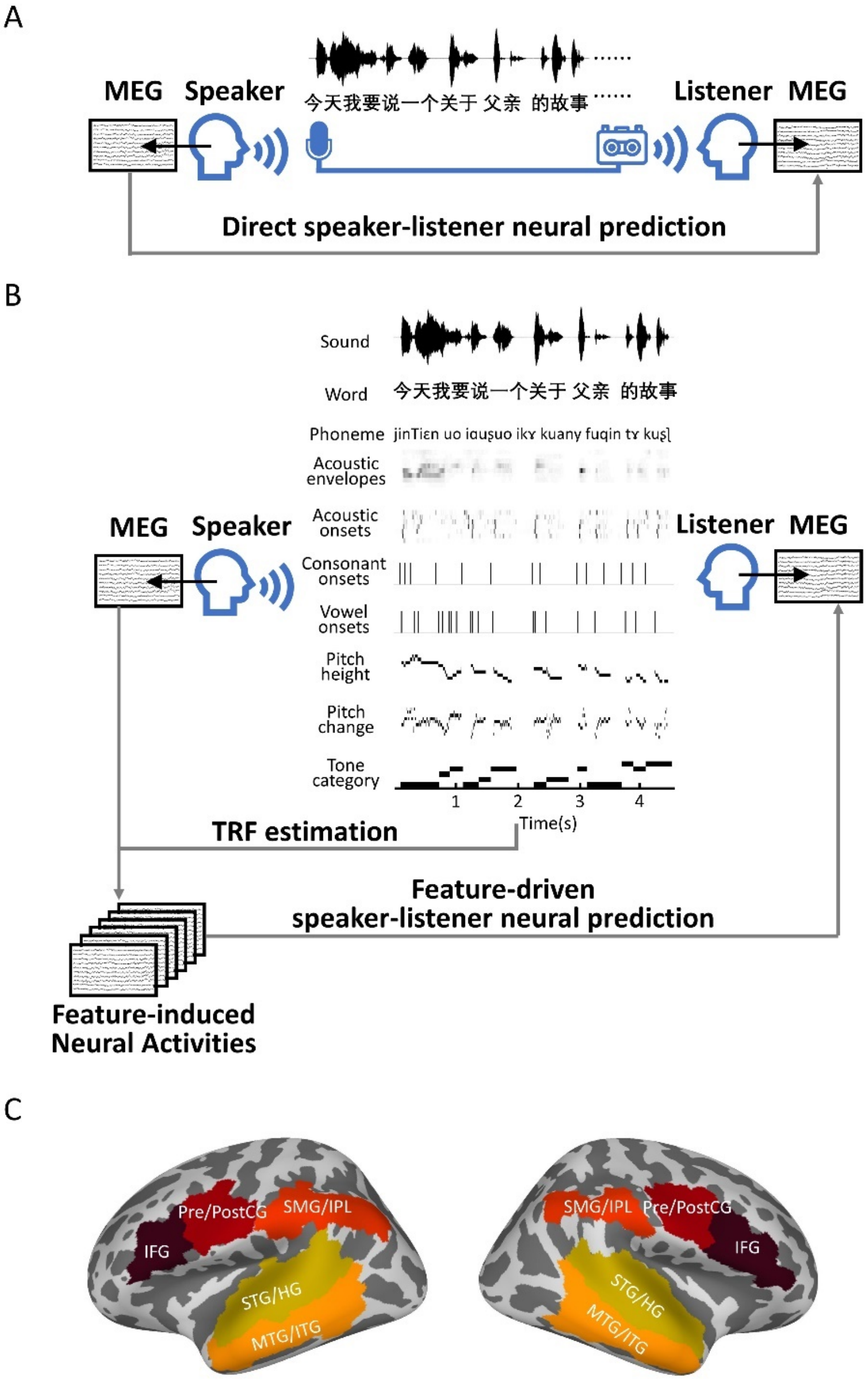

**Figure 1**. Experimental design and analysis framework for speaker-listener neural synchronization (NS). (**A**) Experimental design and direct speaker-listener neural prediction. MEG recordings are obtained from a speaker narrating a story in Mandarin Chinese (example sentence shown: "今天我要说一个关于父亲的故事" / "Today I'm going to tell a story about father"). The recorded audio is then presented to listeners while their brain activity is simultaneously measured with MEG. Neural activity patterns from the speaker are directly used to predict the listener's brain responses. (**B**)

Feature-driven speaker-listener neural prediction. The narrative is analyzed across multiple levels: acoustic envelopes, acoustic onsets, consonant onsets, vowel onsets, pitch height, pitch change, and tone categories. Temporal Response Function (TRF) estimation is applied to quantify how each linguistic feature in the speaker's production predicts neural responses in the listener, resulting in feature-induced NS patterns. (**C**) Brain Regions of Interest (ROIs). The NS analysis examined key cortical ROIs, displayed on bilateral brain surface models. These regions include: IFG (Inferior Frontal Gyrus), Pre/PostCG (Precentral/Postcentral Gyrus), SMG/IPL (Supramarginal Gyrus/Inferior Parietal Lobule), STG/HG (Superior Temporal Gyrus/Heschl's Gyrus), and MTG/ITG (Middle Temporal Gyrus/Inferior Temporal Gyrus). These regions represent the core network of speech production and comprehension.

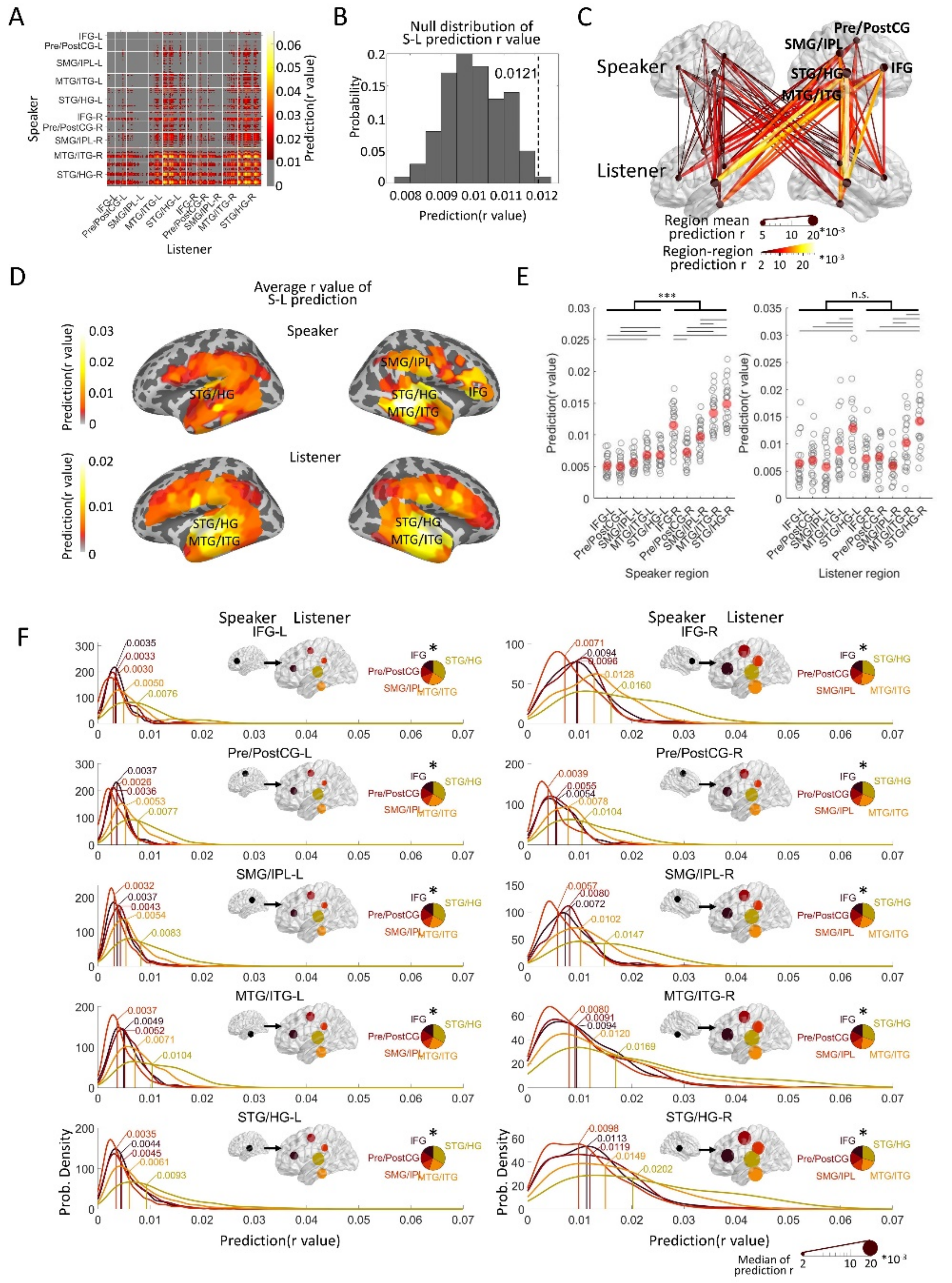

**Figure 2**. Direct neural synchronization between speaker and listener revealed through speaker-to-listener neural predictions. (**A**) The speaker-listener prediction (r values) across all source location pairs, with values below the significance threshold colored in grey. (**B**) The null distribution of prediction r values generated by a permutation test, with a significant threshold set at 0.0121. (**C**) Speaker-listener neural predictions across all significant ROI pairs, where the size and color of the connecting lines indicate the degree of prediction. The size of the dots representing speaker and listener regions reflects the mean predictions across all listener/speaker regions. (**D**) The average inter-brain prediction r values for significant sources, with the upper two panels representing the speaker (157 sources) and the lower two panels representing the listeners (190 sources). (**E**) Compares mean predictions across ROIs and hemispheres for the speaker (left panel) and listener (right panel) individually. Grey circles represent each subject's mean inter-brain prediction; red circles indicate the mean prediction across subjects. The horizontal lines above the circles denote the significance of pairwise post hoc tests. (**F**) The distribution of inter-brain prediction values, with each panel representing the results between one speaker's region and all listener's ROIs. The effects of hemisphere and ROI in NS were visually salient. The median of each distribution is indicated near the corresponding curve. The size of each colored circle on the listener's brain represents the median prediction value from the speaker's region to the listeners' bilateral regions (only the left hemisphere is plotted for clarity), while the pie chart displays their proportional relationships. Significant differences in prediction were observed across all region pairs, as confirmed by a one-way ANOVA ($p < 0.05$, FDR-corrected), denoted by asterisks above the pie charts.

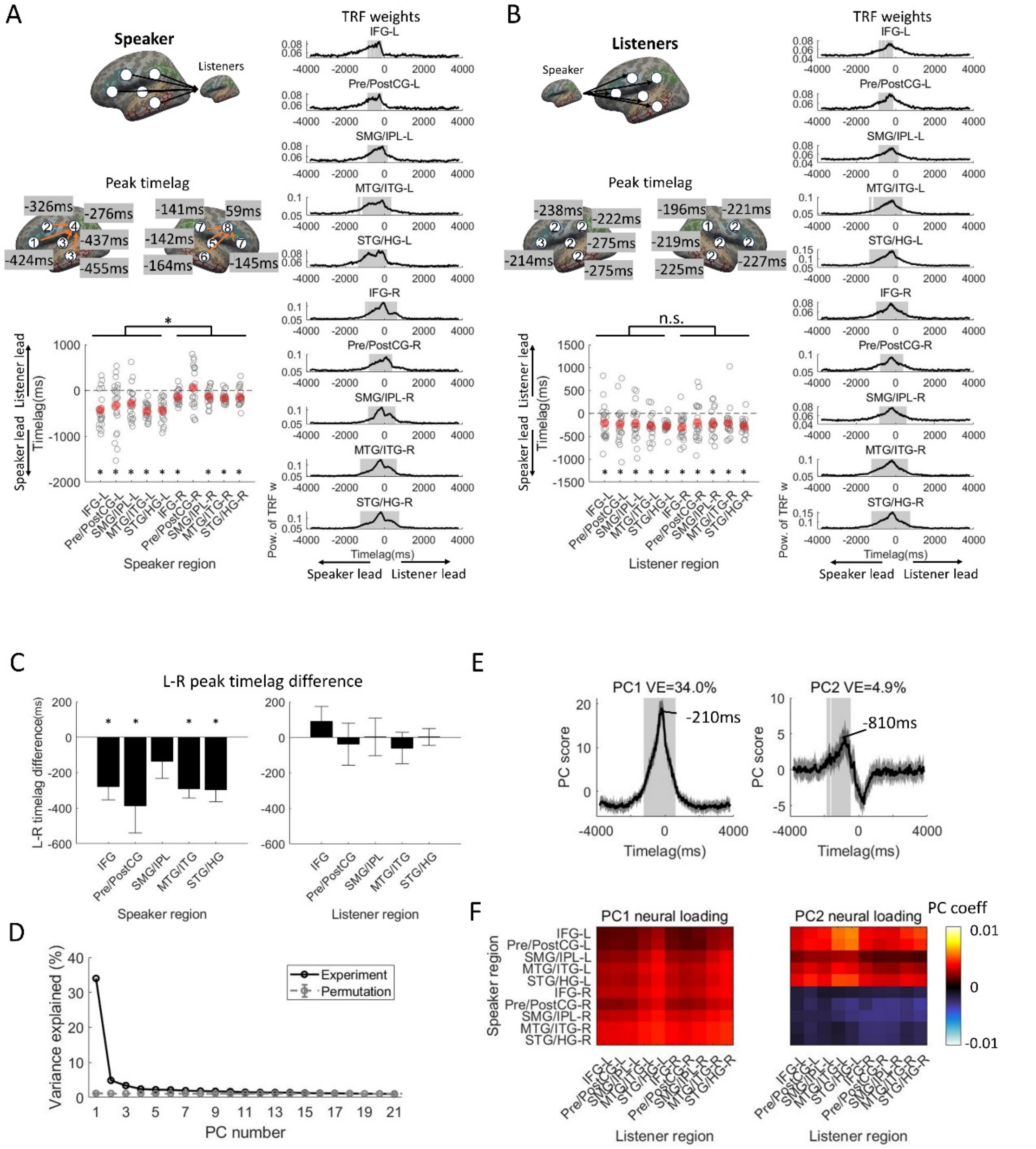

**Figure 3**. Speaker-listener neural synchronization (NS) time-lag patterns. TRF weights across time-lag windows for the speaker (**A**) and the listener (**B**) regions. The upper-left panels display the ROIs for both hemispheres (one hemisphere's graph was displayed for clarity). The middle-left panels illustrate the mean peak time lags across these regions. The numbers within the circles refer to the ranking in time lags. In the lower-left panels, grey circles represent individual subject's time lags, while red circles indicate the average time lags across all subjects. The asterisks below these panels denote the significance of peak latencies compared to zero. The right panels exhibit the mean power-transformed TRF weights across speaker and listener regions, with shaded

areas highlighting periods significantly greater than the threshold estimated via permutation tests. (**C**) Hemisphere differences in peak time lag. Higher negative values refer to more time lags in NS in the left hemisphere compared to the right hemisphere. Asterisks indicate the significance of differences relative to zero; error bar = s.e.m. (**D**) Principal component analysis (PCA) and the variance explained by principal components (PCs), with black lines representing results from the actual experimental data and grey lines indicating permutation results. (**E**) Displays the PC amplitudes (mean ± SEM) for the first (left panel) and second (right panel) significant PCs, with shaded areas signifying time-lag periods that were significantly greater than zero. The peak time lags were highlighted for the PCs. (**F**) Neural loadings of the first (left panel) and second (right panel) PCs across all the speaker-listener ROI pairs. Y-axis = speaker regions; x-axis = listener regions; higher PC loadings refer to more contribution to the PC.

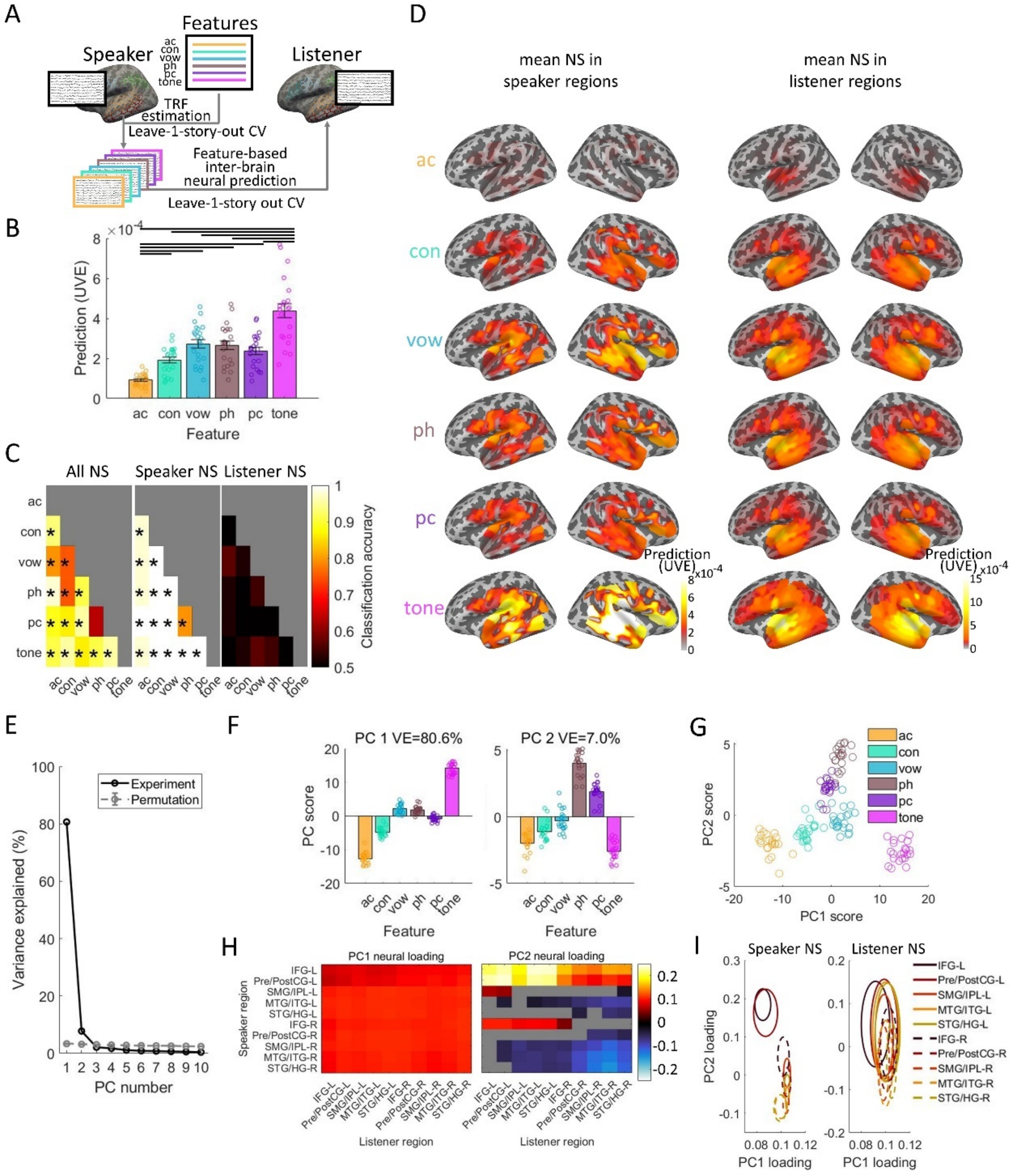

**Figure 4**. Feature-driven speaker-listener neural synchronization. (**A**) Schematic illustration of the feature-based neural encoding modeling procedure for speaker-to-listener neural prediction. Different colors of the lines and box refer to different features (from acoustics to tone category). (**B**) The unique contribution of each feature in inter-brain prediction, i.e., unique variance explained (UVE [$R^2$]) of each feature averaged across speaker-listener region pairs (mean ± SEM). Colored circles depict neural prediction for each speaker-listener pair, while the horizontal lines above indicate the significance of pairwise comparisons. (**C**) Feature classification accuracies based on their inter-brain NS patterns. Asterisks denote the feature pairs (e.g., consonant vs. tone

category) with significant distinct NS patterns. A leave-one-pair-out cross-validation procedure was used with a regularized linear discriminant analysis for the classification. (**D**) feature-driven speaker-to-listener neural predictions for the speaker (left panel) and listener regions mapping on the surface brains. (**E**) Two principal components (PCs) explained over 87% of the total variances of the inter-brain prediction patterns, with black and gray lines representing results from real MEG and shuffled data, respectively. (**F**) PC scores of the two significant PC across six features. The tone category and two pitch features showed the highest weights in contributing to the two PCs, respectively. (**G**) The two PCs created a 2-dimensional neural space that can best distinguish the six features, where acoustics, pitch, and tone category features are located in three corners, while segmental units are located in between. Each colored circle denotes the PC score of each speaker-listener pair. (**H**) Neural loadings for the first (left panel) and second (right panel) PCs across all speaker-listener region pairs. The region pairs with loadings not significantly different from chance were shaded in gray. (**I**) Distributions of the speaker (left panel) and listener (right panel) ROIs in the 2-dimensional neural loading space created by the two significant PCs. The center of each ellipse represents the mean neural loading of each ROI, and the two axes illustrate the standard deviation across the dimensions of the two PCs. (ac: acoustic; con: consonant; vow: vowel; ph: pitch height; pc: pitch change; tone: tone category).

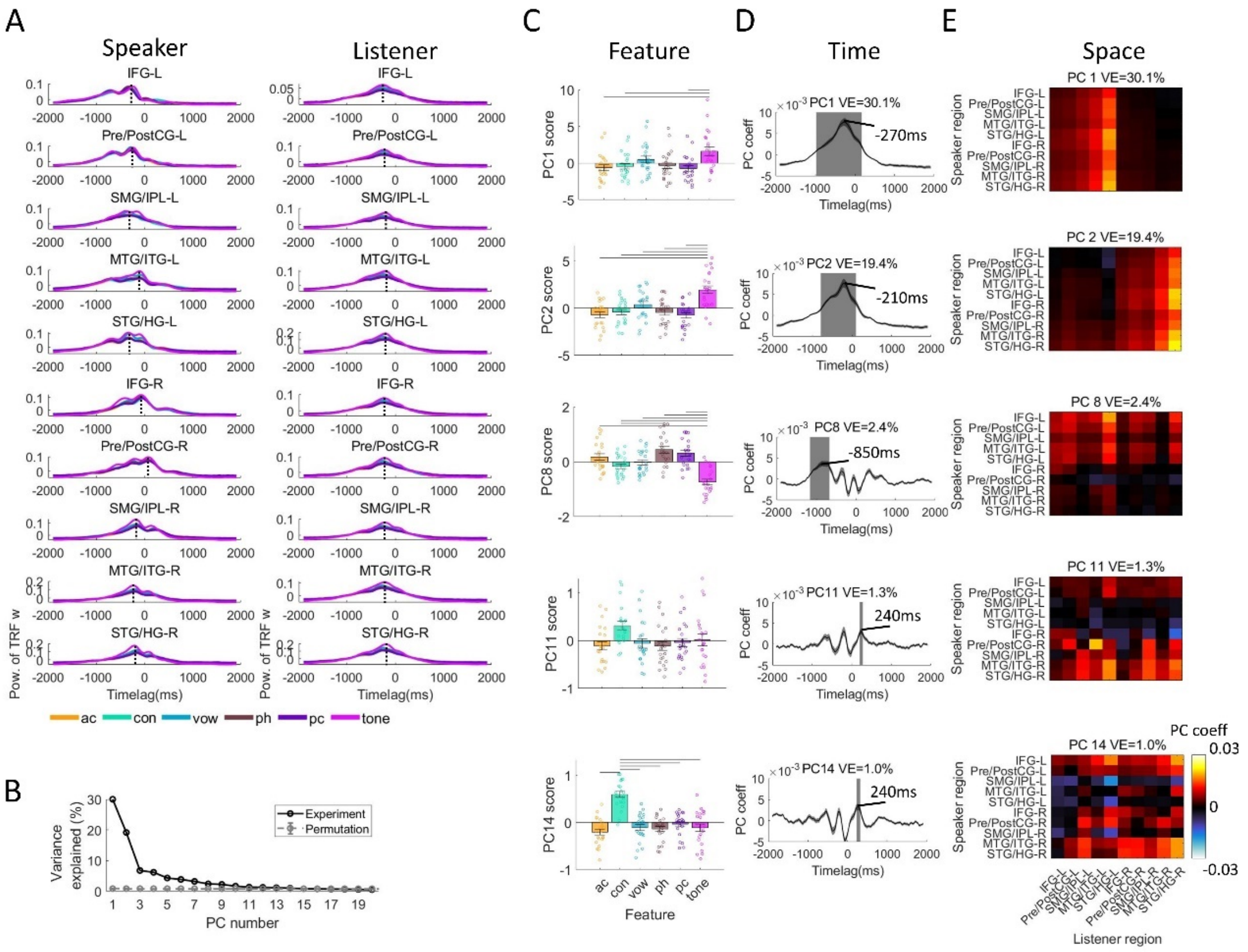

**Figure 5**. Spatiotemporal Patterns of Feature-Driven Speaker-Listener Neural Synchronization. (**A**) Normalized power of temporal response function (TRF) weights for six features (different colors of the lines) across regions (mean ± SEM) of the speaker (left panels) and the listeners (right panels). The dashed vertical lines indicate the peak time lags of the mean of six weights. (**B**) Variance explained by principal components (PCs) to the raw spatiotemporal NS patterns. Black and gray lines represent PCs derived from the actual MEG and shuffled data, respectively. The highest 14 PCs were significantly better than chance, while only five of them showed significant differences in PC scores between features (one-way ANOVA, $p < 0.05$, FDR-corrected). (**C**) PC scores across six features for the five PCs (PC 1, PC 2, PC 4, PC 8, and PC 11). Horizontal lines indicate significant differences between features as determined by post hoc t-tests ($p < 0.05$, FDR-corrected). (**D**) Temporal latency patterns (mean ± SEM) of the five PCs, calculated by averaging the PC loadings across all speaker-listener region pairs. Shaded areas represent time periods where TRF weights are significantly greater than zero (t-tests, $p < 0.05$, FDR-corrected). (**E**) Spatial NS patterns across speaker-listener region pairs for each of the five PCs. These NS patterns were derived from the average of PC loadings within the significant time range shown in panel D.

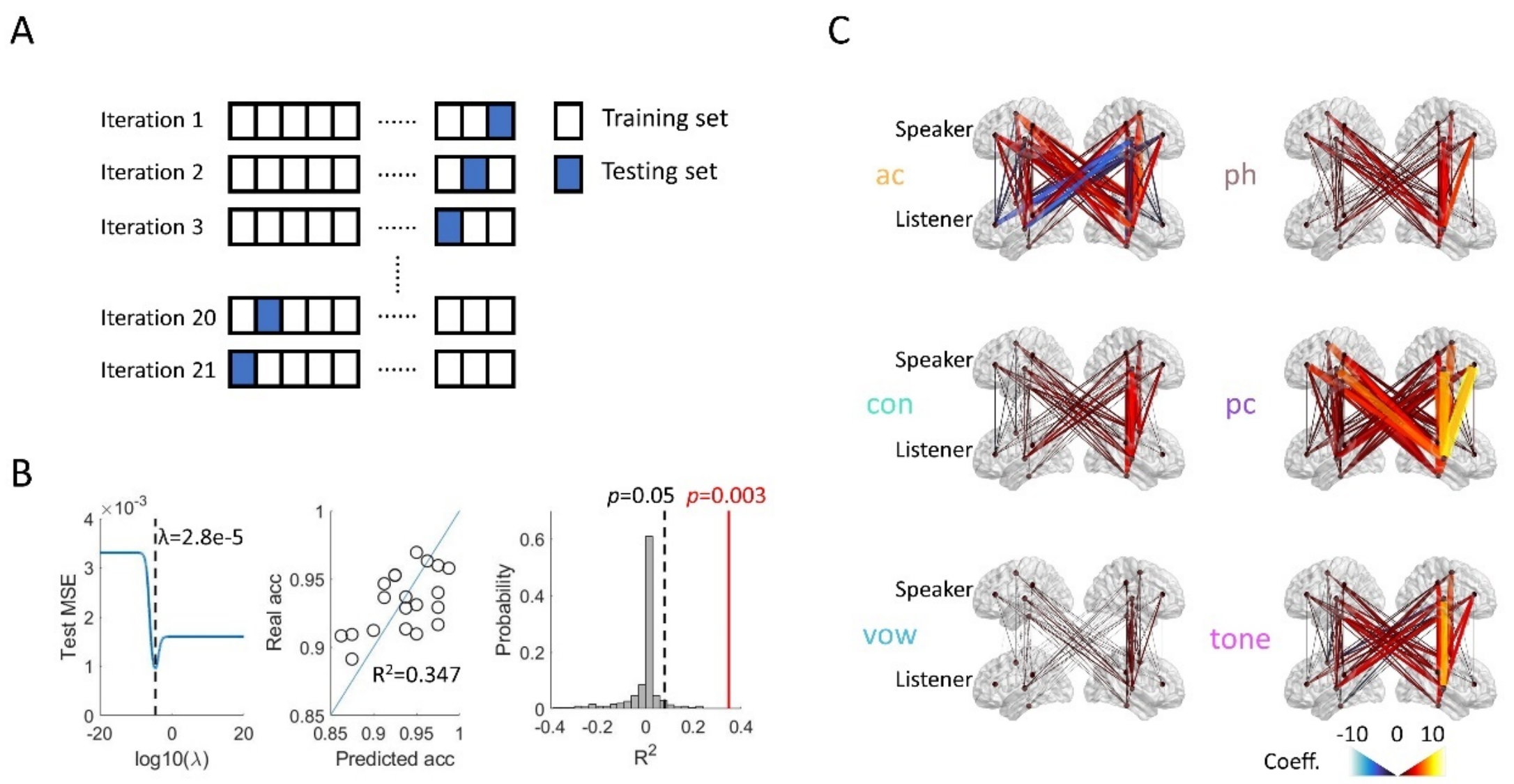

**Figure 6**. Relationship between strength of feature-driven NS and behavior performance. (**A**) Leave-one-out cross-validation procedure of prediction modeling. (**B**) Prediction performance of feature-driven NS patterns. In the left panel, the curve indicates the relationship between test MSE in cross-validation and the logarithm of lambda values. The optimal lambda is 2.8e-5. In the middle panel, each dot represents the relationship between actual behavior performance (the accuracy of answers after listening to stories) and the predicted one of each subject (MSE=0.00095). In the right panel, the grey histogram represents the null distribution of $R^2$ values. The vertical dashed line indicates the 95th percentile of the null distribution (0.079). The red vertical line indicates the real $R^2$ value (0.347). (**C**) Coefficients of feature-driven NS of speaker-listener region pairs. Each panel represents the result of each feature. A larger line indicates a greater contribution to the prediction. Only significant interbrain connections are displayed, corrected $p < 0.05$.

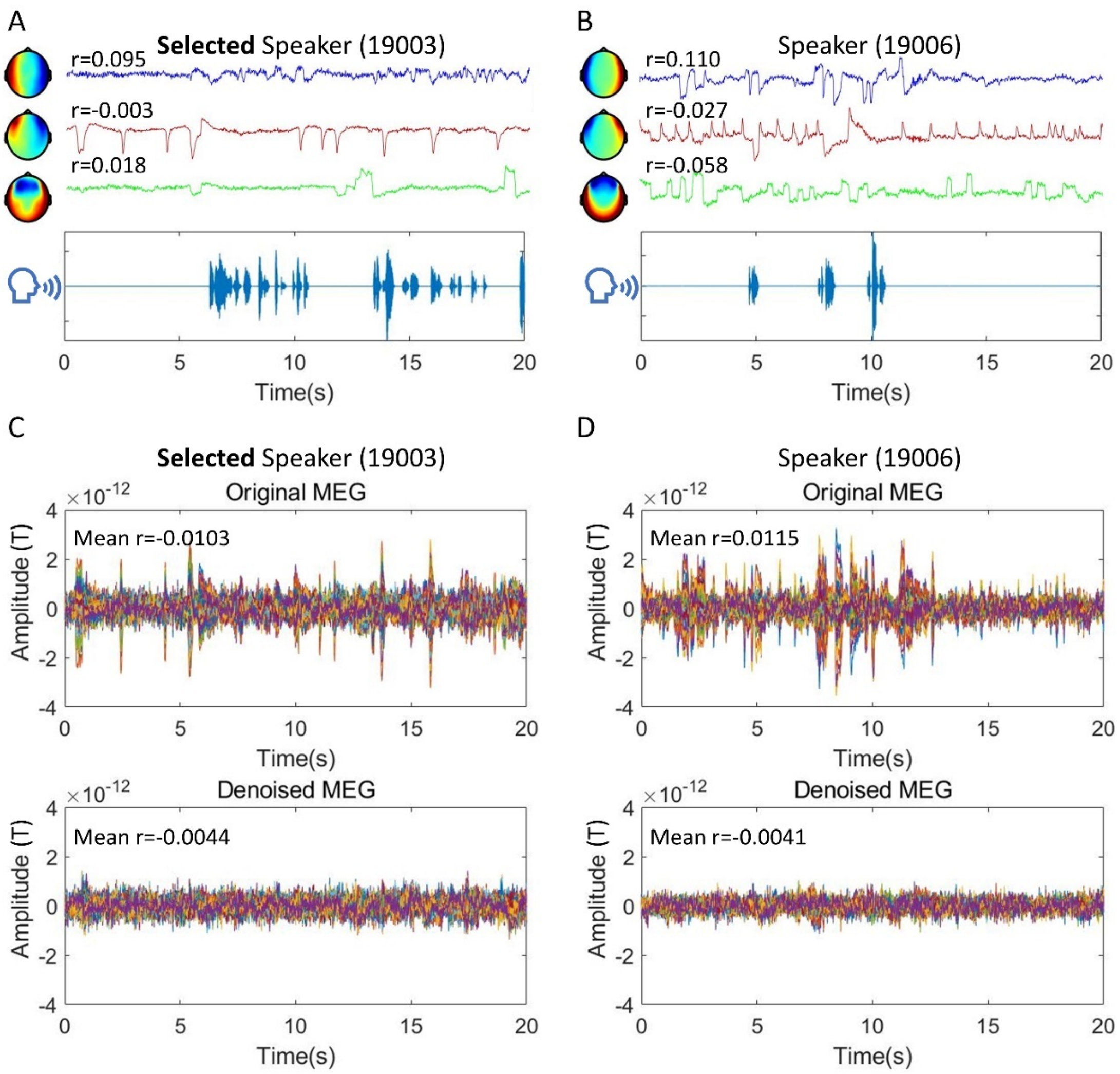

**Figure S1**. MEG signal qualities for two representative speakers with high signal-noise-ratio (SNR). Speaker 19003 was selected for the analyses reported in this study. (**A**) and (**B**) Independent components (ICs) related to muscle movements (top), blinks (middle), and eye movements (bottom) were correctly identified with ICA. The sound waves of the selected speaker (19003, SNR = 8.67 dB) in (A) and another speaker (19006, SNR = 7.12 dB) in (B) are shown along with the ICs. Pearson's correlation (r) between each IC and the sound envelope is indicated near the IC topographies (magnetometers only). (**C**) and (**D**) Original (top panels, bandpass-filtered: 0.3–45 Hz) and denoised (bottom panels, with ICs related to blinks, eye movements, and muscle movements removed) MEG signals are presented for the selected speaker (19003) in (C) and speaker 19006 in (D). Mean Pearson's r values with the sound envelope across all 102 magnetometers were significantly decreased after artifact removal.

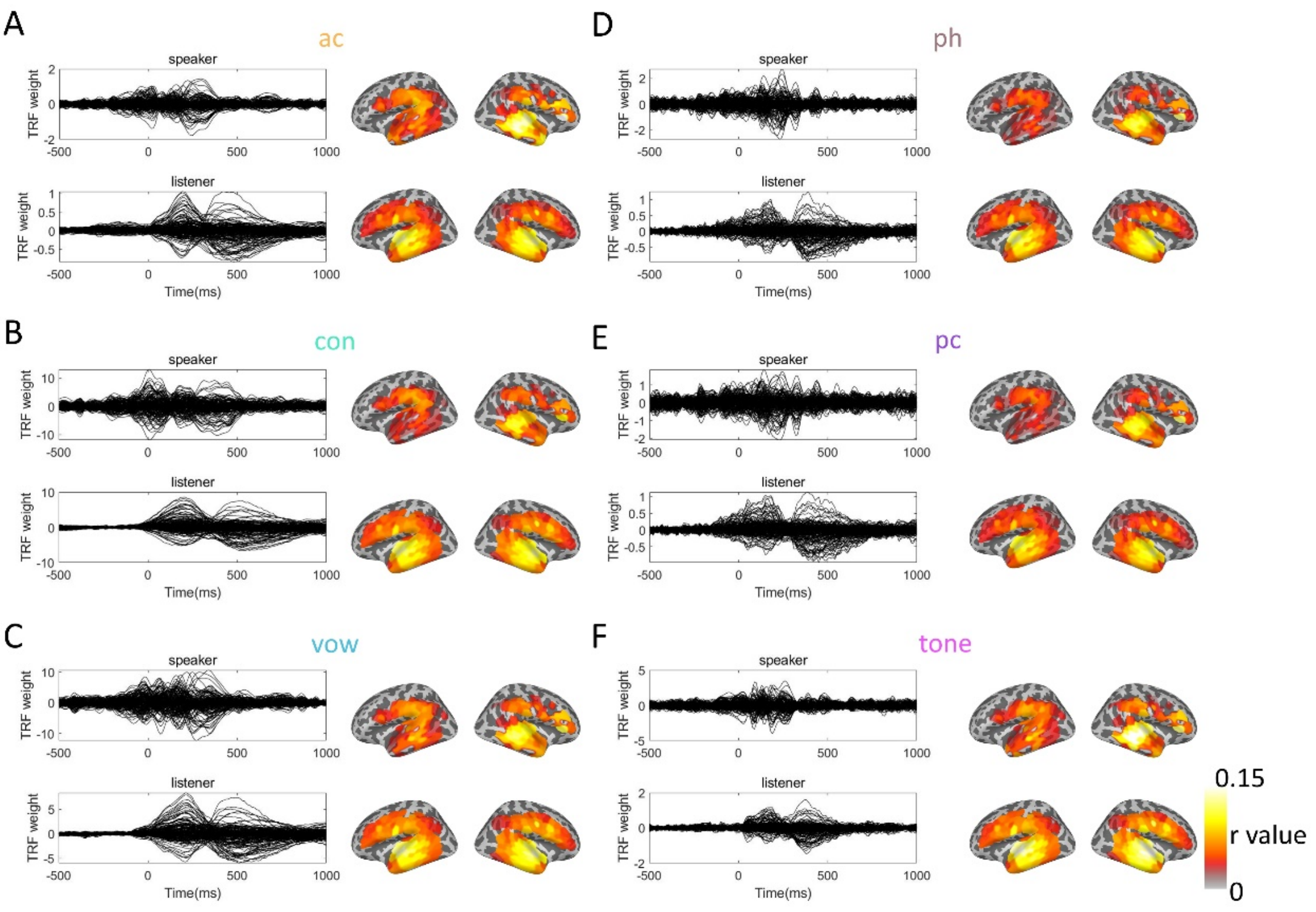

**Figure S2**. Feature-based temporal response functions (TRFs) for speakers and listeners. (**A** to **F**) Each panel presents results for various linguistic features: (A) acoustic (ac), (B) consonants (con), (C) vowels (vow), (D) phonemes (ph), (E) pitch contours (pc), and (F) tone. Left panels: Temporal response function (TRF) weights for the speaker (upper plots) and listener (lower plots) over time, with time 0 marking the onset of the feature. TRF weights represent the neural encoding associated with each feature. Right panels: Distribution of correlation coefficients (r values) for both the speaker (upper brain maps) and listener (lower brain maps) across cortical regions of interest. The color scale illustrates the strength of the correlation (ranging from 0 to 0.15). These findings underscore the differences in feature-specific neural responses between speakers and listeners.

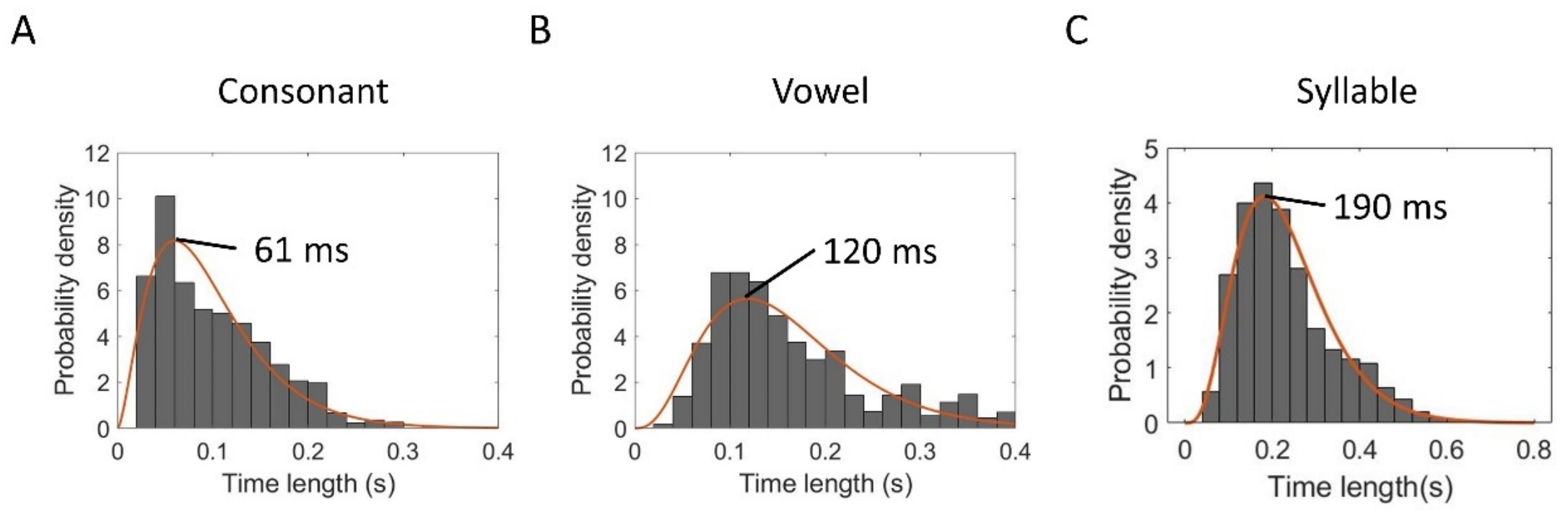

**Figure S3**. Speech production rates of the speaker for consonant, vowel and syllable (also tone). (**A**) and (**B**) The distribution of consonant and vowel durations in the speech stimuli, with a peak around 61 ms and 120 ms, respectively. Consonants and vowels were labeled by trained annotators. (**C**) The distribution of syllable durations in the speech stimuli, with a peak around 190 ms. Syllables were labeled based on the time length of Mandarin characters, corresponding to their spoken duration. The histograms represent the probability density of syllable and phrase durations, with an overlaid curve showing the gamma-fitted distribution. These results illustrate the temporal structure of speech stimuli at different linguistic levels.

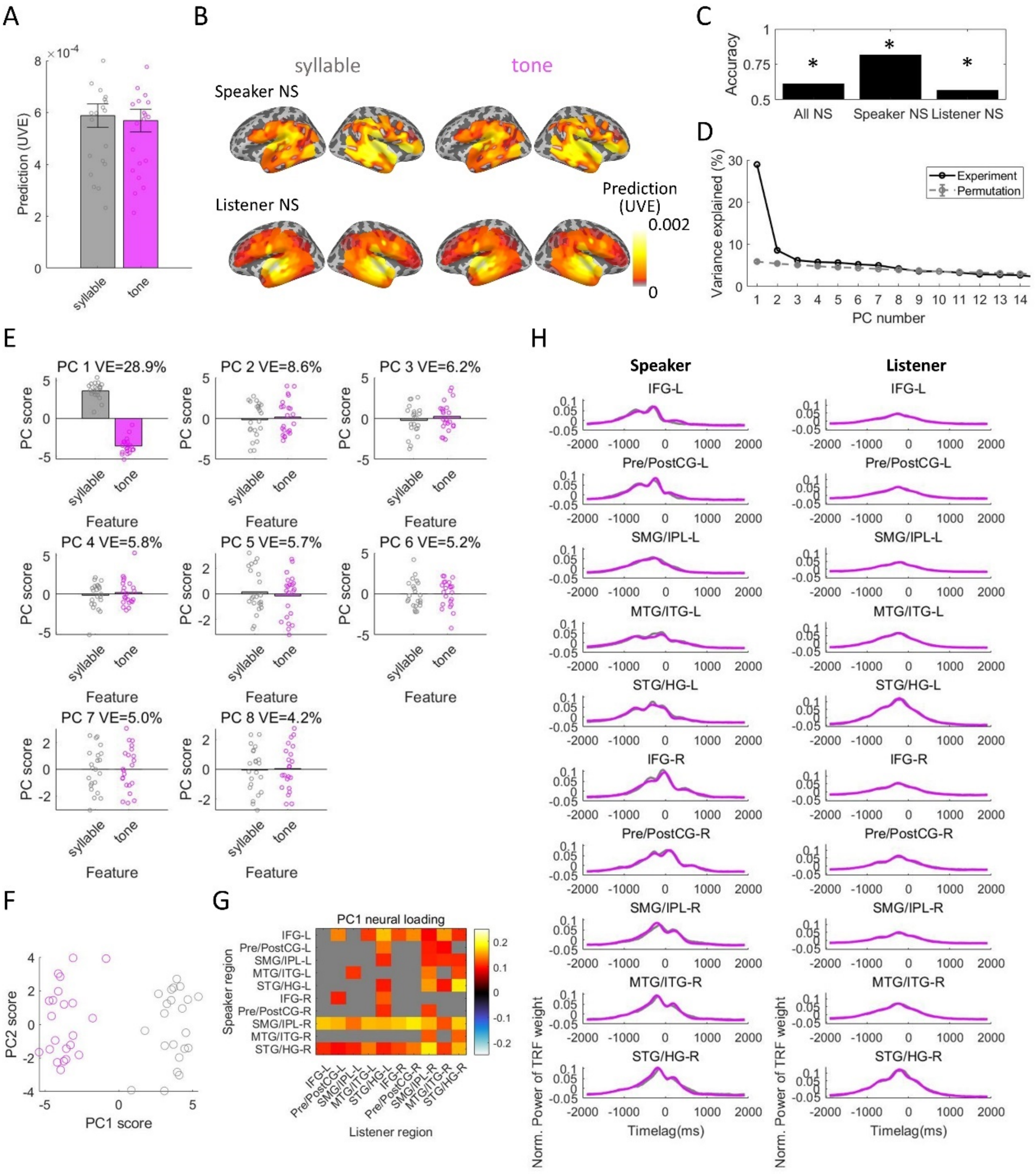

**Figure S4**. Feature-driven speaker-listener neural synchronization (NS) for syllable and tone categories. (**A**) The unique contribution of syllable and tone categories in NS (i.e., interbrain neural prediction). NS indicates the unique variance explained (UVE [R2]), averaged across speaker-listener region pairs (mean ± SEM). Colored circles represent neural predictions for each speaker-listener pair. The two features contributed similarly to NS. (**B**) Feature-driven speaker-to-listener neural predictions for the speaker (upper panels) and listener (lower panels) regions as they map onto the surface brains. (**C**) The classification accuracies of syllable and tone categories rely on their inter-brain NS patterns, indicating that the two features underlie distinct NS spatial patterns.

Asterisks indicate that classification accuracy is significantly better than chance level (permutation test, $p < 0.05$). A leave-one-pair-out cross-validation procedure was performed using regularized linear discriminant analysis for this classification. (**D**) The variance explained by principal components (PCs) corresponding to the NS spatial patterns. Black and gray lines represent PCs derived from the actual MEG and shuffled data, respectively. The top 8 PCs were significantly better than chance. (**E**) PC scores across the two features for the 8 PCs, with the first PC showing significant differences in scores between features (two-sample *t*-tests, $p < 0.05$, FDR-corrected). (**F**) Illustrates that the first two PCs create a 2-dimensional neural space that best differentiates between syllable and tone categories, with each colored circle indicating the PC score for each speaker-listener pair. (**G**) Neural loadings for the first PC across all speaker-listener region pairs, with pairs having loadings not significantly different from chance shaded in gray. (**H**) Normalized power of temporal response function (TRF) weights for the two features (represented by different colored lines) across regions (mean ± SEM) for the speaker (left panels) and listeners (right panels).

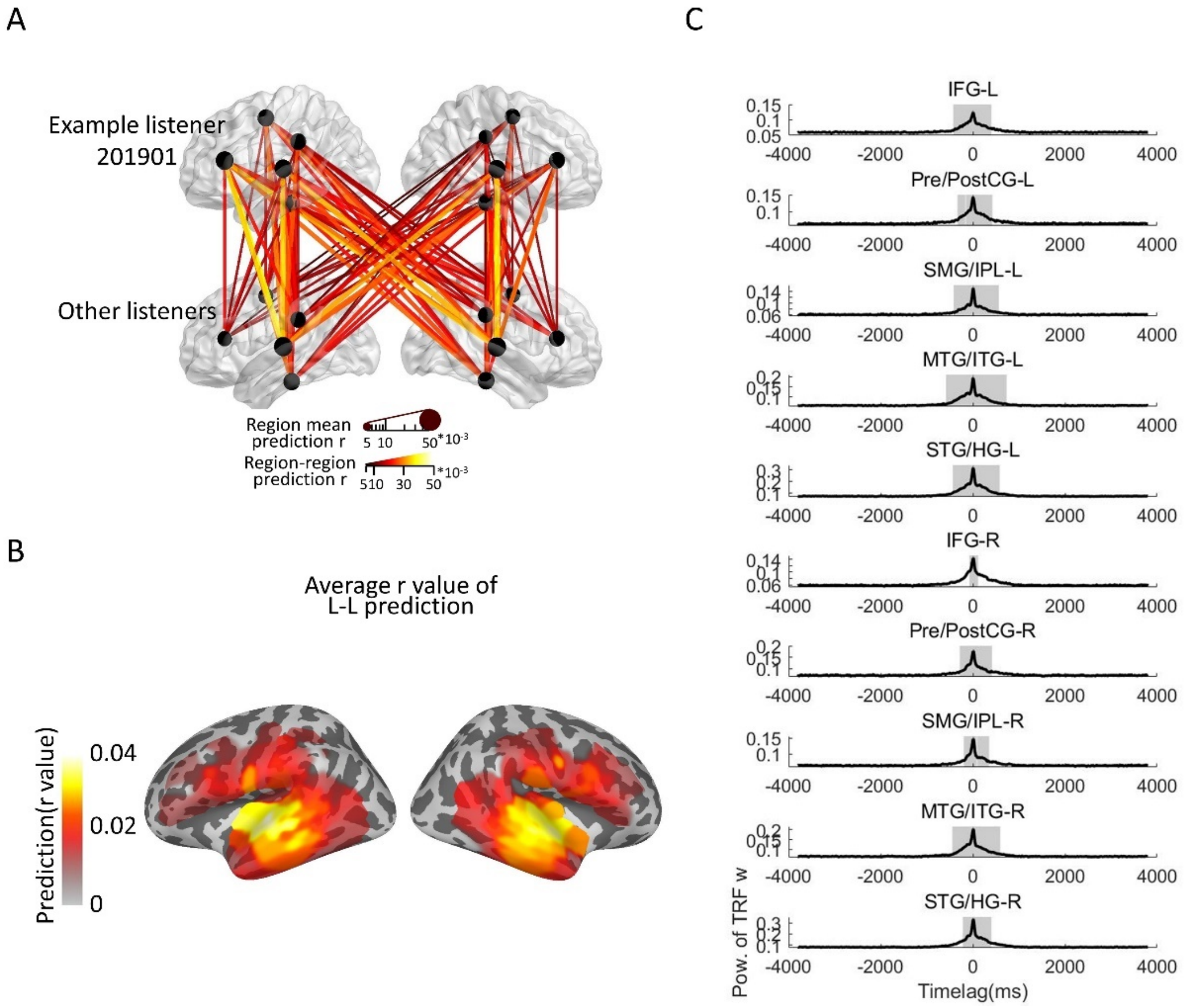

**Figure S5**. Listener-listener neural synchronization (NS) patterns. (**A**) Neural synchronizations between one representative listener and other listeners across all significant ROI pairs, where the size and color of the connecting lines indicate the degree of prediction. The size of regions reflects the mean predictions across all region pairs. Only significant regional pairs are included (permutation test, $p < 0.05$). (**B**) The average interbrain prediction $r$ values across all cortical regions. (**C**) The mean power-transformed TRF weights, with shaded areas highlighting periods significantly greater than the threshold estimated via permutation tests. There is no time lag in NS between listeners.

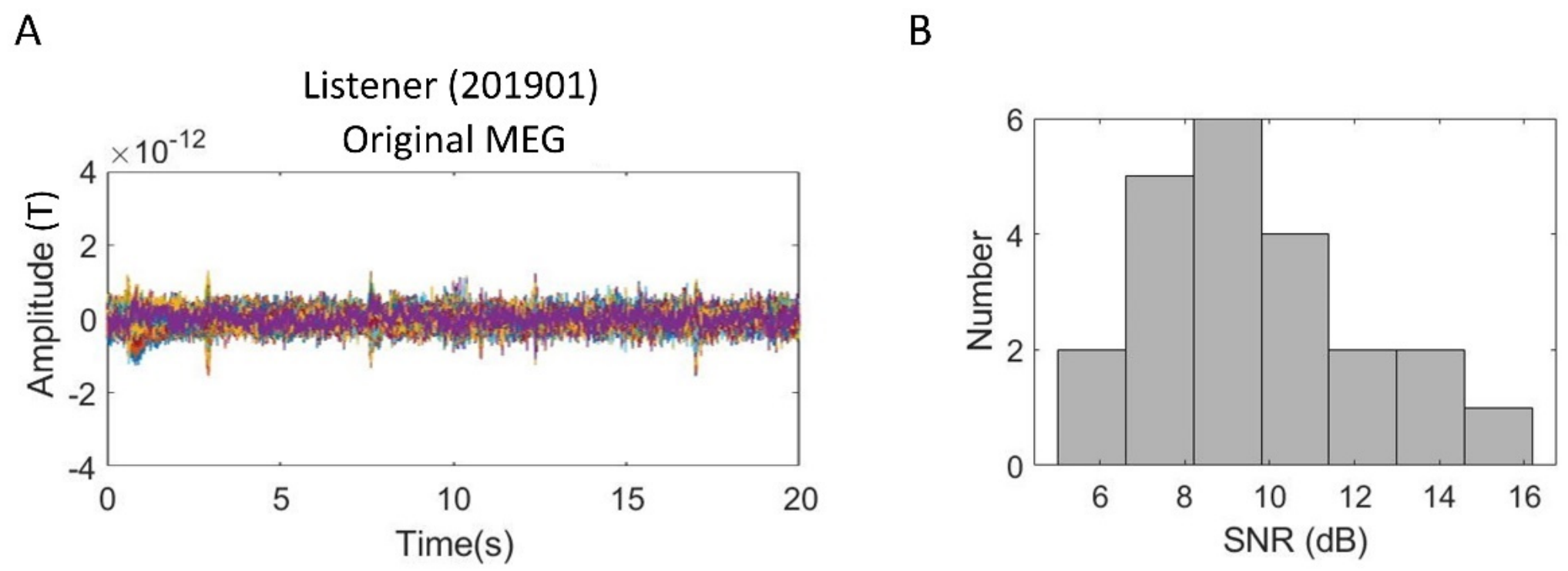

**Figure S6**. MEG signal qualities for 22 listeners. (**A**) Raw MEG signals (bandpass-filtered: 0.3-45 Hz) of one representative listener. All 102 magnetometers are shown. (**B**) Signal-to-noise ratios (SNR) of MEG for the 22 listeners in the listening experiment. The mean is 9.50 dB, and the range is from 5.29 dB to 16.08 dB.

| | Acoustic | Consonant | Vowel | Pitch Height | Pitch Change | Tone Category |
|---|---|---|---|---|---|---|
| $R^2$ | -0.065 | 0.184 | 0.258 | 0.215 | 0.154 | 0.186 |
| *MSE* | 1.55e-3 | 1.18e-3 | 1.07e-3 | 1.14e-3 | 1.23e-3 | 1.18e-3 |

**Table S1**. Prediction performance of listeners' comprehension scores with feature-driven NS patterns.

| | 19002 | 19003 | 19004 | 19006 | 19007 | 19008 |
|---|---|---|---|---|---|---|
| SNR (dB) | 3.20 | 8.67 | 4.41 | 7.12 | 6.49 | 3.59 |

**Table S2**. Signal-to-noise ratios (SNR) of MEG for the six speakers in the production experiment. The audio recordings and MEG data of subject 19003 were chosen based on audio quality, speech production fluency, and MEG SNR. The MEG SNR of subject 19003 (8.67 dB) is similar to those of the listeners (Fig. S6).